\documentclass[preprint,journal]{vgtc}       % preprint (journal style)

\ifpdf%                                % if we use pdflatex
  \pdfoutput=1\relax                   % create PDFs from pdfLaTeX
  \usepackage{graphicx}                % allow us to embed graphics files
  \DeclareGraphicsExtensions{.pdf,.png,.jpg,.jpeg} % for pdflatex we expect .pdf, .png, or .jpg files
\else%                                 % else we use pure latex
  \usepackage{graphicx}                % allow us to embed graphics files
  \DeclareGraphicsExtensions{.eps}     % for pure latex we expect eps files
\fi%

\graphicspath{{figures/}{pictures/}{images/}{./}} % where to search for the images

\usepackage{microtype}                 % use micro-typography (slightly more compact, better to read)
\PassOptionsToPackage{warn}{textcomp}  % to address font issues with \textrightarrow
\usepackage{textcomp}                  % use better special symbols
\usepackage{mathptmx}                  % use matching math font
\usepackage{times}                     % we use Times as the main font
\usepackage{cite}                      % needed to automatically sort the references
\usepackage{tabu}                      % only used for the table example
\usepackage{booktabs}                  % only used for the table example
\usepackage{color}
\newcommand{\revise}[2]{\textcolor{black}{#2}}

\onlineid{1158}

\vgtccategory{Research}
\vgtcpapertype{algorithm/technique}

\title{FibAR: Embedding Optical Fibers in 3D Printed Objects for Active Markers in Dynamic Projection Mapping}

\author{Daiki Tone, Daisuke Iwai, \textit{Member, IEEE}, Shinsaku Hiura, \textit{Member, IEEE}, and Kosuke Sato, \textit{Member, IEEE}}
\authorfooter{
\item
 Daiki Tone and Kosuke Sato are with Osaka University
\item
 Daisuke Iwai is with Osaka University and JST, PRESTO. E-mail: daisuke.iwai@sys.es.osaka-u.ac.jp.
\item
 Shinsaku Hiura is with University of Hyogo.
}

\shortauthortitle{Tone \MakeLowercase{\textit{et al.}}: FibAR: Embedding Optical Fibers in 3D Printed Objects for Active Markers in Dynamic Projection Mapping}
\abstract{%
This paper presents a novel active marker for dynamic projection mapping (PM) that emits a temporal blinking pattern of infrared (IR) light representing its ID.
We used a multi-material three dimensional (3D) printer to fabricate a projection object with optical fibers that can guide IR light from LEDs attached on the bottom of the object.
The aperture of an optical fiber is typically very small; thus, it is \revise{rarely noticed by}{unnoticeable to} human observers under projection and can be placed on a strongly curved part of a projection surface.
In addition, the working range of our system can be larger than previous marker-based methods as the blinking patterns can theoretically be recognized by a camera placed at a wide range of distances from markers.
We propose an automatic marker placement algorithm to spread multiple active markers over the surface of a projection object such that its pose can be robustly estimated using captured images from arbitrary directions.
We also propose an optimization framework for determining the routes of the optical fibers in such a way that collisions of the fibers can be avoided while minimizing the loss of light intensity in the fibers.
Through experiments conducted using three fabricated objects containing strongly curved surfaces, we confirmed that the proposed method can achieve accurate dynamic PMs in a significantly wide working range.
} % end of abstract


\CCScatlist{ % not used in journal version
 \CCScat{K.6.1}{Management of Computing and Information Systems}%
{Project and People Management}{Life Cycle};
 \CCScat{K.7.m}{The Computing Profession}{Miscellaneous}{Ethics}
}

\teaser{
  \centering
  \includegraphics[width=1.0\hsize]{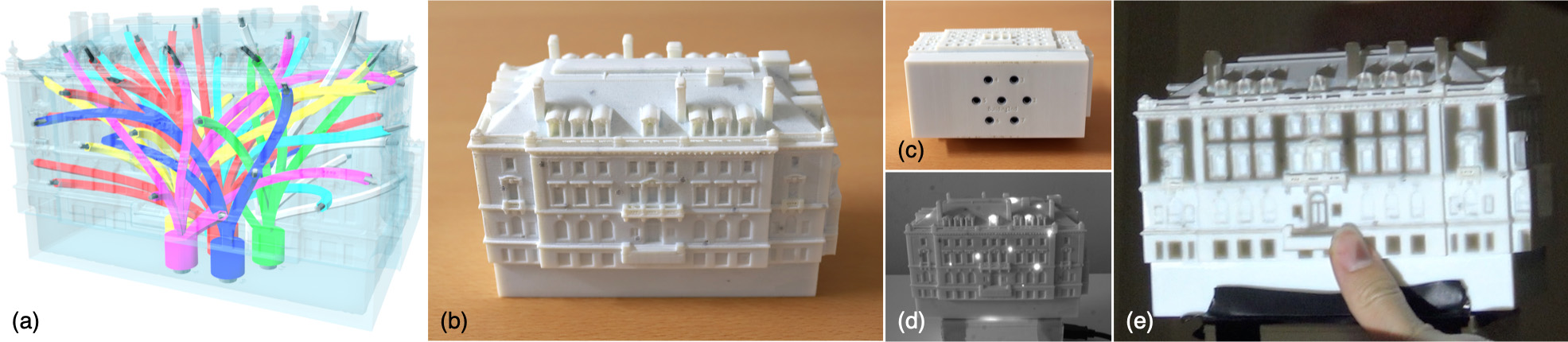}
  \caption{Overview of our proposed system. (a) The internal structure of a projection object. Optical fibers with the same color are connected to the same IR LED attached to the bottom of the object. (b) The projection object fabricated from a multi-material 3D printer. (c) There are seven holes on the bottom of the object, into which IR LEDs are inserted. (d) A captured IR image shows IR light emit from the tip of each fiber, which is transmitted through the fiber from one of the LEDs. (e) A dynamic projection mapping result.}
	\label{fig:teaser}
}

\vgtcinsertpkg

\begin{document}


\firstsection{Introduction}

\maketitle

Projection mapping (PM), also known as spatial augmented reality (SAR) or projection-based AR, can seamlessly merge physical and virtual worlds via projection onto real surfaces \cite{Bimber:2005:SAR}.
It has been integrated and applied in many fields such as education \cite{8172039}, vehicle design \cite{menk2011visualisation,takezawa2019vr}, art creation \cite{flagg2006projector,rivers2012sculpting}, daily life support (e.g., searching everyday objects \cite{iwai2006limpid,iwai2011document}), virtual restoration of historical objects \cite{aliaga2008virtual}, and entertainment (e.g., games \cite{jones2013} and theme parks \cite{mine2012}).
While static surfaces were typically used in these established systems, dynamic PM involving a moving projection surface is still not widely available.
One of the major technical challenges is the geometric registration of a projector to an arbitrarily moving surface as if projected textures are stuck onto it \cite{Grundhoefer:2018:CGF}.

The visual marker-based approach is a robust and computationally efficient solution.
However, there is a unique technical issue inherent in PM that does not have to be considered in typical video see-through ARs, i.e., markers on a projection surface need to be \revise{imperceptible}{unnoticeable} under projection.
Researchers have made efforts to tackle this by proposing several techniques such as drawing markers using a special ink that is visible only in the near-infrared (IR) spectrum \cite{narita2017dynamic} and by diminishing visible markers using a radiometric compensation technique \cite{asayama2015diminishable,asayama2018fabricating}.
%However, current invisible inks are still slightly visible, and the compensation requires several feedback loops between projection and capturing to diminish the markers. In addition,
However, these systems apply spatial patterns in representing the IDs of markers which results in a tradeoff between the marker size and working range.
The markers need to be large enough to be identified by a camera.
As a result, it is necessary to place a marker on a relatively planar surface.
If it is placed on a strongly uneven or curved surface\revise{ such as the pleats and arms of Stanford Lucy \cite{stanfordlucy}}{}, a part of the marker would easily be occluded from the camera, resulting in the failure of ID recognition.
Therefore, the shape of the projection surface is heavily restricted.
Although small markers relax the restriction, they significantly limit the working range, i.e., the systems work only when the distance between the camera and the projection object is short enough to correctly recognize the markers.

In this paper, we propose the use of an active marker emitting a temporal blinking pattern of IR light (which represents its ID), that is not susceptible to the previous tradeoff. % emitted from an end of an optical fiber that is embedded in a projection object.
%The blinking pattern represents the ID of each marker.
A na\"{i}ve implementation would be to embed LEDs in a projection object at all the marker positions.
However, doing so requires tedious manual work to install multiple LEDs under the surface and to connect them to a driving circuit via wires that also need to be manually installed in the object.
On the other hand, by leveraging a recent advancement of the multi-material 3D printing technology, we can apply another strategy that requires almost no manual works for marker installation.
Specifically, a user only needs to attach a base component (consisting of near IR LEDs, a circuit, and a battery) at the bottom of the projection object.
The IR light from the LEDs can be guided through optical fibers to the surface of the projection object. %, from which blinking patterns are emitted.
%The fibers guide the IR light from an LED to the surface of the projection object.
The object can be printed out from a multi-material 3D printer that can automatically embed optical fibers in the object.
The aperture of an optical fiber is typically very small; thus, it is \revise{rarely noticed by}{unnoticeable to} human observers under projection and can be placed on a non-planar part of a projection surface.
In addition, the working range of the proposed system can be larger than the previous marker-based methods because the blinking patterns can theoretically be recognized by a camera placed at a wide range of distances from markers.

\begin{figure}[t]
  \includegraphics[width=0.98\hsize]{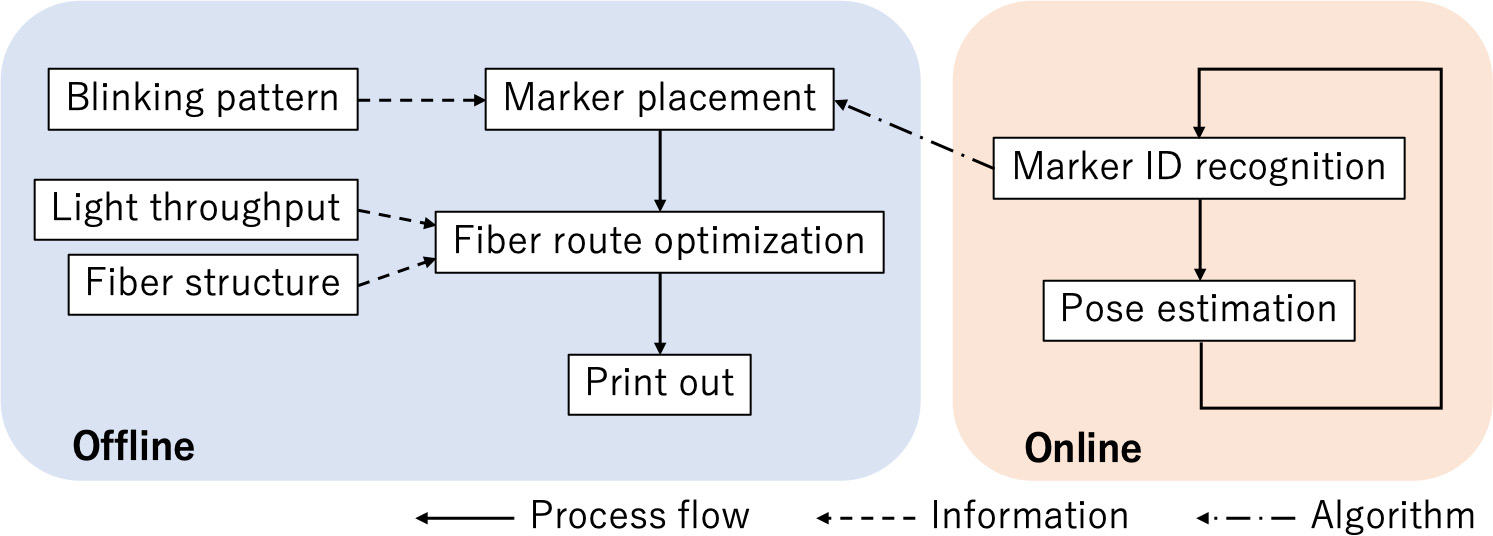}
  \caption{\revise{}{The overall system diagram.}}
  \label{fig:diagram}
\end{figure}

In the proposed method, the placement of each marker on an uneven surface needs to be carefully determined to avoid occlusion.
To this end, we propose an automatic marker placement algorithm for spreading multiple active markers over the surface of a projection object such that its pose can be robustly estimated from captured images obtained from arbitrary directions.
Our algorithm determines the marker placement under the condition that multiple markers share the same blinking pattern and are connected to the same IR LED.
If we simply assign a unique blinking pattern to each active marker, the number of IR LEDs and that of markers need to be identical.
However, the number of markers is relatively large for the use case we considered (e.g., more than 40 in our experiment), and a large number of LEDs cannot be attached to the bottom surface of the object due to its limited space.
Also, long bit depths are needed to assign unique blinking patterns to the markers.
The length of the pattern does not matter as long as tracking of the object is successful, however, it requires a long time to recover from a tracking failure.
Thus, we assign the same blinking patterns to multiple markers to reduce the number of IR LEDs.
%Because each marker should stay relative to the camera to correctly decode its pattern while blinking a whole binary sequence, a long sequence decreases the recognition performance when either of the marker or camera moves.
Once marker placement is decided, the next thing to do is to compute the routes of optical fibers from each LED to the corresponding markers on the surface.
We developed an optimization framework for determining the routes of the optical fibers in such a way that collisions of the fibers can be avoided and the loss of light intensity in the fibers is minimized.
\revise{}{Figure \ref{fig:diagram} shows the overall system diagram.}

In this study, we present the technical details of our automatic marker placement and optimal fiber routing algorithms.
Afterward, we discuss the experiments we conducted using fabricated projection objects to evaluate the tracking accuracy and working range of a prototype system.
To summarize, this paper makes the following three prime contributions:
\begin{itemize}
\item We introduce a novel active AR marker for dynamic PM applications based on optical fibers that are automatically embedded in a projection object, by leveraging the recent multi-material capability in 3D printing technology.
\item We propose an automatic marker placement algorithm to avoid marker occlusions on an uneven surface for robust pose estimation under a condition in which multiple markers share the same binary pattern.
\item We optimize the routes of optical fibers from each LED to the corresponding multiple markers to avoid collision of the fibers and to maximize the intensities of the guided IR lights.
\end{itemize}

%%%%%%%%%%%%%%%%%%%%%%%%%%%%%%%%%%%%%%%%%%%%%%%%%%
\section{Related Works}
\label{sec:relatedworks}
%%%%%%%%%%%%%%%%%%%%%%%%%%%%%%%%%%%%%%%%%%%%%%%%%%

Bandyopadhyay et al. proposed the first prototype of a dynamic PM in which a user draws on a handheld object using projected imagery \cite{bandyopadhyay2001dynamic}.
Since then, dynamic PM has been explored mainly in the research context of user interface \cite{karitsuka2003wearable,cao2007multi,willis2011sidebyside}.
Although many interaction techniques were proposed, accurate geometric registration was out of focus in these projects.
It is only in recent time that researchers have started to focus on manipulating the appearance (e.g., texture or BRDF) of a moving surface rather than just projecting GUI widgets or pictures.
To this end, accurate geometric registration becomes an indispensable technical challenge.
In the subsequent section, we discuss prior works on dynamic PM in the latter category.
% \cite{Grundhoefer:2018:CGF}

Previously presented geometric registration methods in dynamic PM mainly applied computer vision techniques, and fall into two groups: marker-less and marker-based approaches.
The marker-less approach suits the context of PM very well because it does not need to attach markers on a projection surface, which significantly affect projected appearances.
A projector's six-degree-of-freedom (6DOF) transform relative to a rigid-body projection object was successfully estimated by matching its original model with measured information such as color \cite{resch2014sticky}, edges \cite{Morikubo:2018:TPM:3275476.3275494}, and depth images \cite{siegl2015real}.
%Resch et al. applied an RGB camera to track a 3D rigid body surface by matching a predicted projection result and a captured image, followed by a sparse ICP (iterative closest point) algorithm .
%Siegl et al. employed a depth camera to track their projectors 6 degree-of-freedom (DOF) transform relative to the object based on a ICP technique \cite{siegl2015real}.
%The marker-less methods typically worked well only in specific situations.
However, these method cannot be applied to projection surfaces having either symmetrical structures (e.g., flat, cylindrical, and spherical surfaces) or periodic shapes (e.g., wavy surfaces) because their 6DOF poses cannot be estimated uniquely.
%It is also still difficult to independently track multiple objects, particularly when they come close each other.
In contrast to the works dealing with rigid-body objects, Bermano et al. proposed a method to manipulate the appearance of a human face, a non-rigid, and deformable surface, by applying a commercially available face tracker \cite{bermano2017makeup}.
However, this method worked only for a very specific surface (i.e., face).
Recently, Miyashita et al. proposed the application of high-speed cameras measuring the surface normal of a projection object and to manipulate the apparent surface material by projecting directionally varying colors based on target BRDFs \cite{Miyashita:2018:MPM:3272127.3275045}.
Although this method works for various types of surfaces including fluid, it does not support the attachment of texture onto a moving object.

Generally speaking, marker-based methods work more robustly in situations where the marker-less methods do not work well.
Researchers, as well as media artists, created impressive dynamic PM experiences by leveraging the tremendous advancements in motion capture technologies \cite{inori}.
However, it is unfortunate that we may see motion capture markers on a projected surface in an installation, which significantly degrade a sense of immersion in the experience.
Therefore, one of the main technical issues in the marker-based approach is to make markers \revise{imperceptible for}{unnoticeable to} human observers, while being detectable by a camera.
Researchers proposed drawing AR markers on a surface with IR inks, which absorb incident light in the IR spectrum \cite{punpongsanon2015projection,narita2017dynamic}.
The markers have very low contrast in the visible spectrum, although they are still visible for human observers.
Asayama et al. proposed a framework of visually diminishing markers by applying a closed-loop radiometric compensation technique \cite{asayama2015diminishable,asayama2018fabricating}.
However, their methods require several feedbacks of projection and capturing, which can take longer than a second to converge.
\revise{}{Other researchers tried to hide markers from human observers by embedding spatial holes inside projection objects, which can be detected only by Terahertz \cite{Willis:2013:IFI:2461912.2461936} or IR  \cite{Li:2017:AUP:3126594.3126635} imaging.}
Another technical issue concerns the size of a marker.
Previous techniques applied relatively large markers in representing IDs of markers and increasing the robustness in marker detection \cite{asayama2018fabricating,narita2017dynamic}.
As a result, the markers cannot be placed on a strongly uneven surface that can easily cause occlusion of a part of each marker, resulting in marker detection and recognition failure.
Although small markers can be used to solve this problem, they limit the range of the distance between the camera and the object where the markers in the captured images can be correctly recognized.

In this study, we apply an active marker to solve the above-mentioned problems of the marker-based methods.
Several active marker approaches for AR have been proposed.
The HiBall Tracker is a motion capture system for virtual reality (VR) and AR applications that measures the 6DOF pose of the tracker on which multiple photo-diodes (PDs) are installed for measuring the positions of ceiling-mounted IR LEDs that are sequentially flashed \cite{welch2001high}.
Matsushita et al. firstly proposed the application of blinking light from an IR LED as an AR marker \cite{matsushita2003id}.
They used a customized camera called ID CAM that can decode the ID of an AR marker by itself.
While these systems use binary codes, Naimark and Foxlin proposed the use of amplitude modulation codes that enables decoding of an AR marker's ID without synchronization between LEDs and cameras \cite{naimark2005encoded}.
\revise{}{Recently, the synchronization problem has been solved by phase shifting technique of the blinking pattern \cite{10.1145/3332165.3347884}.}
\revise{}{Mohan et al. proposed Bokode, an LED-based spatial marker \cite{Mohan:2009:BIV:1531326.1531404}. A tiny lens is placed on an object's surface and the marker is embedded behind the lens at its focal distance. A camera focusing at infinity is used to capture the marker.}
Active markers were also realized by an opposite way, where a projector projects per-pixel IDs which are measured by PDs.
Most previous works developed special high-speed projectors to embed binary codes \cite{raskar2007prakash,zhou2014dynamically}.
On the other hand, Kitajima et al. proposed to directly use a normal laser projector and decode the pixel position information from the raster scanning timing \cite{kitajima2017simultaneous}.
Although these previous works worked well, LEDs or PDs with electrical wires had to be manually embedded underneath a projection surface at predefined positions.
This process is cumbersome and causes significant errors in pose estimation of the surfaces.

Recently, the huge potential in 3D printing of optics has been recognized in digital fabrication as well as optics research fields.
For instance, researchers showed the possibility of a multi-material 3D printer for embedding optical fibers in a 3D printed object \cite{willis2012printed}.
We apply this concept to avoid a complex work of embedding active marker LEDs in a projection object.
In particular, we propose to attach the LEDs at the bottom of the projection object and connect them to markers on the surface using printed optical fibers.
Attaching LEDs to the bottom of the object is much easier than embedding them underneath the surface.
In addition, the aperture of an optical fiber is typically very small, and thus, the markers are potentially \revise{rarely perceivable by}{unnoticeable to} human observers.
Furthermore, the blinking patterns from the optical fiber are theoretically detectable from a camera at a wide range of distances.
However, owing to low transparency of currently available clear materials (e.g., Stratasys {\it VeroClear}) used in multi-material 3D printers and their low printing  resolutions, the light throughput of a printed optical fiber significantly reduces according to the increase in the length and curvature of the fiber.
Pereira et al. proposed optimizing the routes of a bunch of printed optical fibers to maximize light throughput \cite{pereira2014computational}.
However, their technique does not consider a complex situation where fibers cross each other in a printed object.
In this paper, we present a novel fiber routing algorithm for handling such a situation to realize reliable active markers.
\revise{}{A closely related previous work proposed a computational tool for determining internal pipe routes in 3D printed objects for various interactive applications \cite{Savage:2014:STA:2642918.2647374}. However, this did not consider the optical fiber routing problem. To the best of our knowledge, we are the first group that tries to optimize the routes of printed fibers.}

%%%%%%%%%%%%%%%%%%%%%%%%%%%%%%%%%%%%%%%%%%%%%%%%%%
\section{Active Markers}

This section discusses our active marker design.
First, we explain the blinking or temporal patterns of the markers.
Second, we present how to recognize markers on a projection surface from captured images.
\revise{The last subsection introduces}{Third, we introduce} our proposed automatic marker placement algorithm that decides the position of each marker on the surface to make its pose estimation robust.
\revise{}{The last subsection explains our computer vision technique to estimate the pose of the projection object from captured markers.}

\subsection{Blinking pattern}
\label{sec:marker:pattern}

To make the system as simple as possible, it is preferable that the system does not have a synchronization circuit between a camera and active markers.
Instead, we embed a synchronization signal into the blinking pattern of IR light.
Particularly, we apply an m-sequence, a pseudorandom binary sequence of $2^m-1$ bits, which is widely used in wireless communication as well as visible light communication \cite{qiu2016let}.
M-sequence has a couple of important properties that make it suitable for our purpose.
First, the autocorrelation is 1 for zero-lag and nearly zero for all other time lags.
Second, the phase of the sequence can be obtained from the latest $m$ bit data.
In this study, we assign the pattern ID (denoted as $u_{ptn}$) of 1 to an m-sequence code.
We assign other pattern IDs ($u_{ptn}=2,3,\ldots,N_{ptn}$) to arbitrary binary codes whose code lengths are $b_{ptn}$ such that $((2^m-1)\ {\rm mod}\ b_{ptn})=0$ and $b_{ptn}\le(2^m-1)/2$.
Namely, these binary codes repeat $(2^m-1)/b_{ptn}$ times in a cycle of the m-sequence code.
To reduce the bit depth of the patterns and decrease the number of LEDs, we assign the same pattern ID to multiple markers.
%All IR LEDs attached to the bottom surface of a projection object are synchronized by a single driving circuit.

The proposed system captures the sequence of blinking patterns of multiple markers using a camera, from which we identify the phase and pattern IDs.
As mentioned above, we can determine the phase of the m-sequence from the latest $m$ bit information of the pattern $u_{ptn}=1$.
However, another code (i.e., $u_{ptn}\ne1$) may correspond to it.
In that case, we analyze at most the latest $2b_{ptn}$ bits and determine the pattern ID as 1 if the cycle of the pattern does not correspond to $b_{ptn}$.
%For example, if we apply $a=4$ (i.e., the code length of the m-sequence is 15) and $L=3$, we need to check the latest 6 bit information at most.
If the pattern ID is 1, then we determine the phase by comparing the obtained bit information with the original m-sequence.
We then determine the pattern IDs of the other patterns based on the phase.

\subsection{Marker ID recognition}
\label{sec:marker:ID}

Markers are placed on surface points of a projection object.
We refer to these surface points as marker points.
As already mentioned, we assign the same pattern ID to multiple markers; thus, the number of pattern ID is smaller than that of marker ID.
%Suppose $N_{mkr}$ represents the number of marker IDs, then, $N_{ptn}\le N_{mkr}$.
We can recognize the marker ID of a marker point from the pattern IDs of that point and the adjacent marker points.
Previous methods applied the same approach \cite{narita2017dynamic,watanabe2017extended}.
They placed markers on grid points, and this geometric constraint made the retrieval of adjacent marker information simple.
However, it is not always possible to place markers on grid points in our proposed method owing to a constraint in optical fiber routing as described in Section \ref{sec:fiber}.
Therefore, we propose a marker ID recognition method that allows for more flexible marker placement.

\begin{figure}[t]
  \includegraphics[width=0.98\hsize]{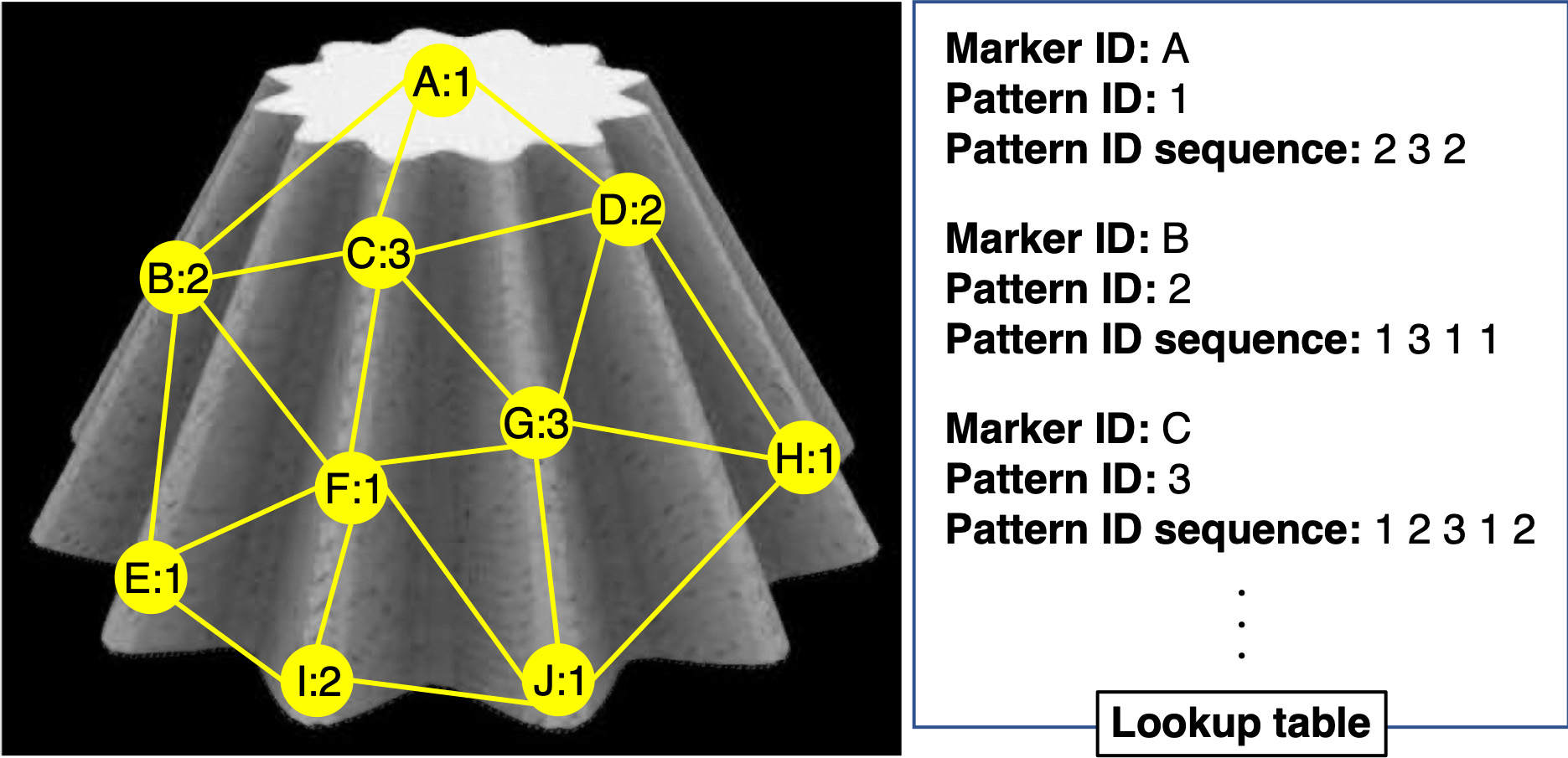}
  \caption{Example of connected adjacent markers overlaid on a simulated image and a lookup table of the markers.}
  \label{fig:database}
\end{figure}

In offline, we build a lookup table, in which each marker point $p_d$ is linked to the corresponding marker ID $u_{mkr}(p_d)$, pattern ID $u_{ptn}(p_d)$, and the sequence of the pattern IDs of adjacent markers ${\bf u}_{ptn}(p_d)$ (Figure \ref{fig:database}).
Considering occlusion of adjacent markers during marker detection, ${\bf u}_{ptn}(p_d)$ is generated based on simulated images that are rendered by capturing the computer graphics model of the projection object from various viewpoints.
The viewpoints are uniformly distributed over a sphere centered at the center of the model.
Particularly, the viewpoints are assigned to the vertices of 4-frequency dodecahedral geodesic sphere in our experiment.
In each rendered image, we connect markers using the Delaunay triangulation algorithm.
Then, for each marker point $p_d$, we trace the directly connected marker points in a clockwise direction to generate the sequence of the pattern IDs of adjacent marker points in the viewpoint.
%We denote the sequence as ${\bf u}_{ptn}(p_d,v)$, which potentially varies among different viewpoints due to occlusions.
The sequence potentially varies among different viewpoints due to occlusions.
Therefore, we select the most frequently occurring sequence and store it in the lookup table as ${\bf u}_{ptn}(p_d)$.
%For each ${\bf u}_{ptn}(p_d,v)$, we compute the sum of Levenshtein distances from the other sequences.
%Then the stored sequence is determined as it takes the minimum.
%Thus,
%
%\begin{eqnarray}
%	\hat{v} = \argmin_v \sum_{\{v'\mid v\ne v'\}}|{\bf u}_{ptn}(p_d,v)-{\bf u}_{ptn}(p_d,v')|_{LD}, \label{eq:marker:1}\\
%	{\bf u}_{ptn}(p_d) = {\bf u}_{ptn}(p_d,\hat{v}), \label{eq:marker:2}
%\end{eqnarray}
%
%where $|\cdot|_{LD}$ represents the Levenshtein distance.

In online, at first, we identify the pattern IDs of marker points in a captured image using the method discussed in Section \ref{sec:marker:pattern}.
Afterward, we identify their marker IDs as follows.
The pattern ID of each captured marker point $p_c$ is represented as $u_{ptn}(p_c)$.
As in the offline simulation, we connect marker points in the captured image using the Delaunay triangulation algorithm.
For each marker point, we obtain the sequence of the pattern IDs of the connected markers in a clockwise direction, denoted as ${\bf u}_{ptn}(p_c)$.
Next, we search in the lookup table for marker points that have the same pattern ID of $u_{ptn}(p_c)$.
Among these marker points in the lookup table, we find the one (denoted as $\hat{p}_d$) corresponding to $p_c$ such that the Levenshtein distance between ${\bf u}_{ptn}(p_c)$ and ${\bf u}_{ptn}(\hat{p}_d)$ is minimum.
%the database for the marker ID of each $p_c$
%At first, we select marker points in the database whose pattern ID $u_{ptn}(p_d)$ is same as $u_{ptn}(p_c)$.
%Then, we find a marker point $\hat{p}_d$ such that the Levenshtein distance between ${\bf u}_{ptn}(p_c)$ and ${\bf u}_{ptn}(p_d)$ takes the minimum.
%
%\begin{equation}\label{eq:marker:3}
%	\hat{p}_d = \argmin_{\{p_d\mid u_{ptn}(p_d)=u_{ptn}(p_c)\}} |{\bf u}_{ptn}(p_c)-{\bf u}_{ptn}(p_d)|_{LD}.
%\end{equation}
%
Finally, we identify $p_c$'s marker ID $u_{mkr}(p_c)$ as $u_{mkr}(\hat{p}_d)$.
%
%\begin{equation}\label{eq:marker:4}
%	u_{mkr}(p_c) = u_{mkr}(\hat{p}_d).
%\end{equation}
%

\subsection{Marker placement}
\label{sec:marker:place}

Markers' locations on a projection surface can significantly affect the estimation performance of the surface pose, and thus, they need to be carefully determined.
Because we estimate the pose by solving PnP (Perspective-n-Point) problem \cite{Lepetit2008}, more than four markers need to be always visible to the camera.
%For example, in an extreme case where markers are placed only on one side of an object, its pose cannot be estimated from a camera observing the other side.
%Therefore, markers should be equally distributed over the surface.
%In previous works \cite{watanabe2017extended,asayama2018fabricating}, marker placement was optimized by maximizing the number of markers visible from various directions.
%However, relatively large markers were applied to represent IDs or increase the detectability; and thus, these methods did not support surfaces having uneven and thin shapes, which cause occlusions of the markers frequently.
Therefore, a straightforward strategy is to maximize the number of markers and equally distribute them over the surface.
However, we need to consider other constraints that are unique to our optical fiber-based markers.
First, due to the internal volume of the optical fibers (see Section \ref{sec:fiber}), the distance between two markers on the surface should be large enough to avoid collision of fibers.
Second, due to the narrow light directivity of the printed fiber, the blinking light from the marker becomes too dark to be detected by a camera when viewed from shallow angles. % with a normally limited dynamic range.
In addition, pattern IDs need to be carefully assigned to the markers so that the marker ID recognition robustly works for various viewing directions.

\begin{figure}[t]
  \includegraphics[width=0.98\hsize]{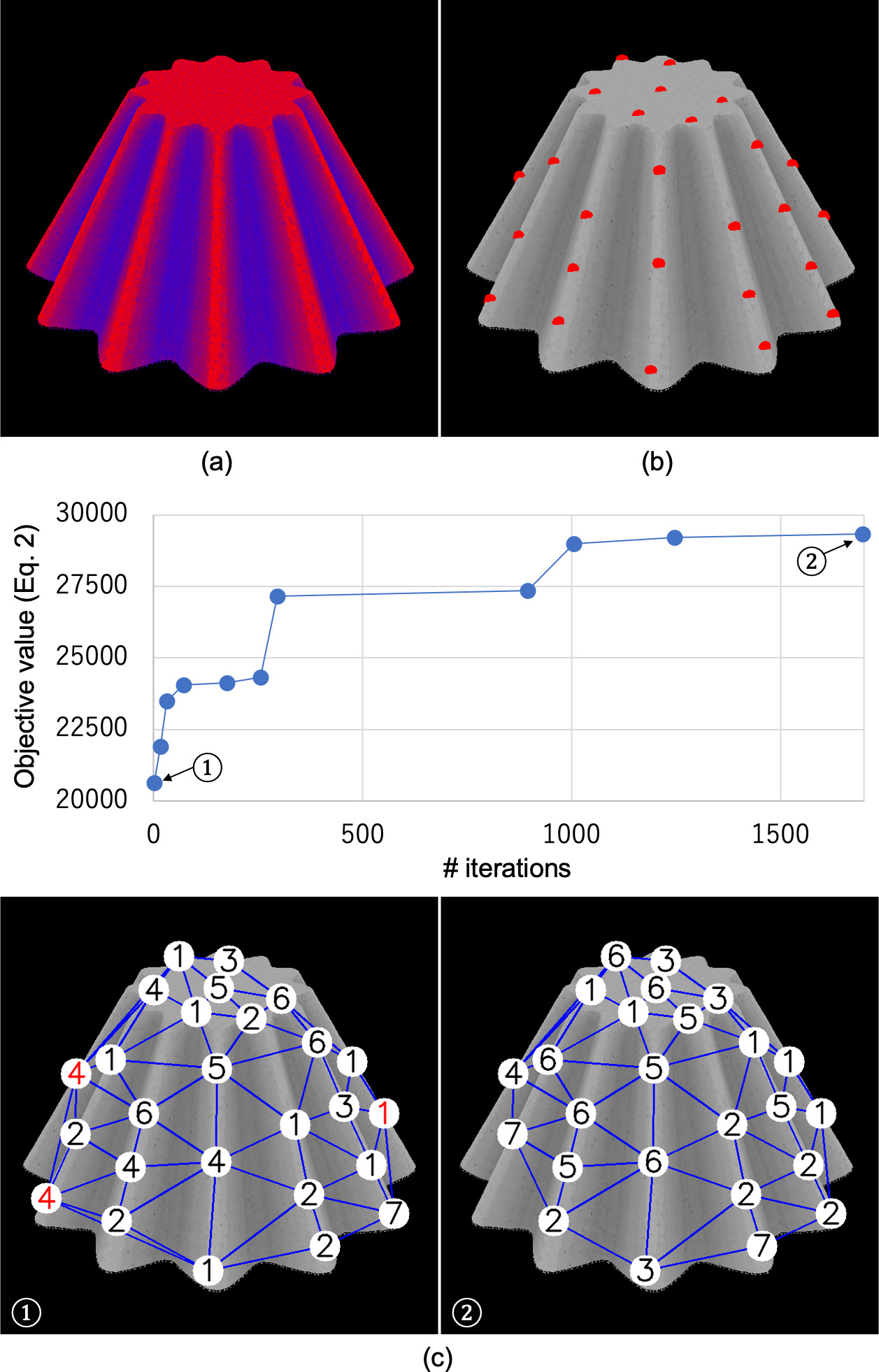}
  \caption{\revise{Initial marker placement for a wavy cone surface: (left) visualization of $s(p_m)$ (red: large, blue: small), (right) initial candidate marker places as red points.}{Marker placement for a wavy cone surface: (a) Visualization of $s(p_m)$ (red: large, blue: small). (b) Initial candidate marker places as red points. (c) Pattern ID assignment by GA (top: the objective value of Equation \ref{eq:marker:6} is improved in the iterations, bottom: marker assignments at the first and last iterations where red numbers indicate that the marker IDs of these markers cannot be recognized in the simulation).}}
  \label{fig:marker_place}
\end{figure}

We propose an automatic marker placement algorithm to spread multiple active markers over the surface of a projection object such that its pose can be estimated from a captured image of the surface by a camera placed at arbitrary directions.
The algorithm assigns each marker to a surface point where marker occlusion and intensity drop-off of IR light do not significantly affect the pose estimation performance of the object.
The algorithm at first determines initial candidate marker points on the surface by evaluating the visibility of IR lights.
Given the 3D model of the projection object, we compute the following value $s(p_m)$ representing the suitability of marker place at each surface point $p_m$:
\begin{equation}\label{eq:marker:5}
	s(p_m) = \sum_{v} vis(p_m,v)\theta(p_m,v),
\end{equation}
where $vis(p_m,v)$ denotes a binary variable representing the visibility of a surface point $p_m$ from a viewpoint $v$.
$vis(p_m,v)$ is 0 when $p_m$ is not visible from $v$, and 1 otherwise.
$\theta(p_m,v)$ represents the angle between the normal vector of $p_m$ and incident vector from $v$ to $p_m$.
%Also note that we assume each projection object has a bottom area where LEDs are embedded and we do not assign markers.
The viewpoints are the same as those used in Section \ref{sec:marker:ID}.
%prepared such that they uniformly distribute over a sphere centered at the center of the object (e.g., one viewpoint each 10 degrees in azimuth and altitude angles respectively in our experiment).
\revise{In addition, w}{We add surface points having locally maximum values of $s(p_m)$ to a candidate list of marker points as follows. First, we store all surface points in the list. Then, we randomly select a point $p_m$, and remove points $p_m'$ around it from the candidate list if $s(p_m')<s(p_m)$ and the distance between these two points is less than a predefined threshold. After repeating this process for all remaining points in the candidate list, we obtain the final candidate list in which surface points having spatially locally maximum values of $s(p_m)$ remain.}
Figure \ref{fig:marker_place}(a, b) shows the visualization of $s(p_m)$ of a wavy cone surface and the initial candidate marker points.

Our algorithm then assigns the optimal pattern ID $u_{ptn}(p_m)$ to each surface point $p_m$ in the candidate list based on their adjacent relationships.
% searches for the optimal pattern IDs for the marker points in the candidate list, based on their adjacent relationships.
In this process, inappropriate marker positions are also discarded from the list.
%We assign a pattern ID $u_{ptn}(p_m)$ to each surface point $p_m$ in the candidate list.
As described above, the same pattern ID is assigned to multiple markers.
The assignment is optimized through a genetic algorithm (GA), in which the array of pattern IDs is a chromosome.
As described in Section \ref{sec:marker:pattern}, $u_{ptn}(p_m)=1$ is for the blinking pattern of m-sequence, and $u_{ptn}(p_m)\ge2$ is for other binary patterns.
In addition, we prepare $u_{ptn}(p_m)=0$ for the case where it turns out that $p_m$ is not suitable for a marker point and discarded in the following optimization.

The optimization is performed based on simulated images that are rendered by capturing the 3DCG model of a projection object from various viewpoints $v$.
The viewpoints are the same as those used in Section \ref{sec:marker:ID}.
Our GA algorithm evaluates each chromosome based on the correctness of marker ID recognition and the visibility of m-sequence markers.
% how the marker IDs are correctly recognized and how many marker points to which the m-sequence pattern is assigned are visible in the images.
Suppose $E_{rcg}(v)$ and $E_{mseq}(v)$ evaluate the former and the latter for each rendered image of viewpoint $v$, respectively, the objective of the GA is
\begin{equation}\label{eq:marker:6}
	{\rm maximize}\ \sum_{v} E_{rcg}(v)E_{mseq}(v).
\end{equation}

The first term, $E_{rcg}(v)$, is computed as follows.
At first, for each marker point $p_m$, we obtain the pattern ID $u_{ptn}(p_m,v)$ and the sequence of adjacent pattern IDs ${\bf u}_{ptn}(p_m,v)$ in the rendered image of viewpoint $v$, which are then used to recognize the marker ID $u_{mkr}(p_m,v)$, by applying the algorithm discussed in Section \ref{sec:marker:ID}.
%first of all, we obtain the sequence of adjacent pattern IDs of $p_m$, i.e., ${\bf u}_{ptn}(p_m)$, by connecting marker points using the Delaunay triangulation algorithm and tracing the marker points directly connected to $p_m$ clockwise.
%For each rendering image from a viewpoint $v$, we recognize the pattern ID of $p_m$, i.e., $u_{ptn}(p_m,v)$, using Equation (\ref{eq:marker:3}).
Subsequently, we compute $E_{rcg}(p_m,v)$ such that it takes the maximum value when the marker ID is correctly recognized and the sequence of adjacent pattern IDs is the same as the most frequently occurring sequence from all the viewpoints, which is denoted as ${\bf u}_{ptn}(p_m)$.
Thus,
\begin{eqnarray}
	\lefteqn{E_{rcg}(p_m,v) =} \\
	&\left\{
	\begin{array}{ll}
		0 & (u_{ptn}(p_m)=0) \\
		k_1-|{\bf u}_{ptn}(p_m,v)-{\bf u}_{ptn}(p_m)|_{LD} & (u_{mkr}(p_m,v)\ {\rm is\ correct}) \\
		-k_1 & ({\rm otherwise})
	\end{array}
	\right. \nonumber
	,
\end{eqnarray}
where $|\cdot|_{LD}$ represents the computation of Levenshtein distance.
$k_1$ is an arbitrary parameter.
A large $k_1$ value ensures that markers are correctly recognized from various viewpoints, while reducing the number of markers (i.e., increasing the number of points of $u_{ptn}(p_m)=0$).
Finally, $E_{rcg}(v)$ is computed as
\begin{equation}
	E_{rcg}(v)=\sum_{p_m}E_{rcg}(p_m,v).
\end{equation}

The second term of Equation (\ref{eq:marker:6}), $E_{mseq}(v)$, evaluates the number of m-sequence markers visible in each rendered image.
We denote the number as $n_{mseq}(v)$.
Theoretically, one m-sequence marker is sufficient to acquire the phase of blinking patterns.
However, due to occlusion and image noise, it is possible that the m-sequence marker is not recognized in a captured image even when it is within the view frustum of a camera.
Suppose the number of m-sequence markers needed to be within the frustum is denoted as $k_2\ (>1)$, the second term is
\begin{equation}
	E_{mseq}(v)=g(\min(k_2,n_{msec}(v))),
\end{equation}
where $g$ can be arbitrary monotonically increasing function.
We found $g(x)=\sqrt{x}+1$ worked well in our experiment.

The GA stops its iterations when the objective value (Equation (\ref{eq:marker:6})) is not improved in the latest 1,000 iterations.
Then, we remove $p_m$ whose pattern ID is 0 from the candidate list of marker points.
We finally apply the $p_m$ in the latest candidate list as marker points and their pattern IDs as $u_{ptn}(p_m)$.
\revise{}{Figure \ref{fig:marker_place}(c) shows the result of the pattern ID assignment by GA.}

\subsection{\revise{}{Pose estimation of projection object from markers}}

\revise{}{To estimate the pose of the projection object, we apply the following computer vision algorithm to the current captured image. First, we binarize the image, extract blobs of the bright pixels, and exclude those of small areas as noise. Then, we compute the bounding boxes (BBs) of the remaining blobs. We track the markers by updating their BBs. If the BB of a blob is overlapped with that of a marker, we regard that the blob belongs to the marker and set the blob's BB as the marker's BB. If the BB of a marker is not updated, we regard that the LED of the marker turns off. If the BB is not updated for more than the period of the m-sequence, we regard that the marker is no longer visible from the camera. For each marker, if the marker ID is already identified in the previous frame, we use it in the current frame. Otherwise, we identify it by the method described in Section \ref{sec:marker:ID}. Then, we estimate the pose of the projection object by solving PnP problem using a RANSAC (RANdom SAmpling Consensus) algorithm. Due to the image noise, some marker IDs are possibly not correctly identified. Therefore, we correct them using the estimated pose.}

%%%%%%%%%%%%%%%%%%%%%%%%%%%%%%%%%%%%%%%%%%%%%%%%%%
\section{3D Printed Optical Fibers}
\label{sec:fiber}
%%%%%%%%%%%%%%%%%%%%%%%%%%%%%%%%%%%%%%%%%%%%%%%%%%

We embed optical fibers inside a projection object, both of which are printed out from a multi-material 3D printer.
The optical fibers connect each IR LED to corresponding marker points on the projection surface, which share the same pattern ID.
We carefully design the fibers as outlined herein.
First, we provide the structure of the optical fiber.
Second, we explain the computational model that computes the light throughput of fiber and determine the parameters of the model through an experiment.
The last two subsections describe our fiber route optimization framework.

\subsection{Structure of printed optical fiber}
\label{sec:fiber:structure}

\begin{figure}[t]
  \centering
  \includegraphics[width=0.8\hsize]{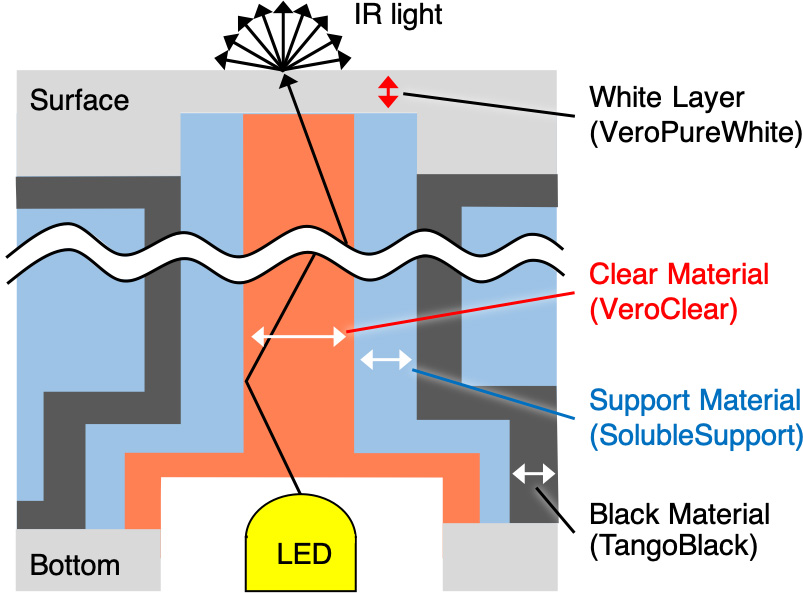}
  \caption{Optical fiber design.}
  \label{fig:fiber}
\end{figure}

Figure \ref{fig:fiber} shows the internal structure of an optical fiber printed out from a multi-material 3D printer.
In this paper, optical fibers are printed out from Stratasys Objet260 Connex3 which can print a 3D object with 3 materials and support material.
We employ white ({\it VeroPureWhite}, RGD837), black ({\it TangoBlack}, FLX973), clear ({\it VeroClear}, RGD810), and support ({\it SolubleSupport}, SUP706) materials.
A typical optical fiber consists of a core surrounded by a transparent cladding material with a lower index of refraction.
According to previous works, we apply the clear and support materials as the core and cladding of our optical fiber, respectively \cite{willis2012printed,pereira2014computational}.
To avoid crosstalk of IR light leaked from fibers inside a projection object, we cover them with a thin layer of the black material.
We cover the object surface with a thin layer of the white material to increase both light diffusion of IR lights and the invisibility of the markers for human observers.

%The performance of the optical fiber is dependent on the thickness of each material.
The diameter of the core should be large to increase the light throughput.
%At the same time, in our context, the diameter should also be small enough so that it is not perceivable by human observers.
On the other hand, thin optical fibers make it possible to embed large number of markers.
To balance them, we determine the diameter of the core as 1.75 mm through several trials and errors.
We also find that the total internal reflection occurs when the thickness of the cladding is more than 0.5 mm.
The leak of IR light does not occur when the thickness of the black layer is more than 0.5 mm.
Therefore, the diameter of the core, the thickness of the cladding, and that of the black layer are set as 1.75 mm, 0.5 mm, and 0.5 mm, respectively.
Namely, the thickness of our printed optical fiber is 3.75 mm.
The projection object is filled with the support material and covered with a 0.4 mm layer of the white material.
The thickness of the white material is also determined through several trials and errors.
\revise{}{Note that previous researchers reported that the core of much smaller diameter worked for their printed optical fibers \cite{willis2012printed}. We consider that this inconsistency comes from the following two factors. First, the light wavelengths are different between our method and \cite{willis2012printed}. Second, our method needs to recognize blinking patterns by a camera, which requires higher contrast fibers than \cite{willis2012printed}.}

\subsection{Light throughput}
\label{sec:fiber:throughput}

To optimize the routes of optical fibers inside the projection object, it is necessary to know the light throughput of the fiber.
According to Pereira et al. \cite{pereira2014computational}, the light throughput $T_f$ of a route $f$ can be modeled using the Lambert-Beer law as follows:
\begin{eqnarray}
	T_f=\exp(-\int_fa),\\
	a=c_1+c_2\exp(-c_3r),
\end{eqnarray}
where $a$ and $r$ represent the absorption coefficient and the curvature radius of the fiber, respectively.
$c_1$, $c_2$, and $c_3$ are arbitrary coefficients.

The light throughput is generally characterized by the luminance of an emitted light from the end of the fiber.
However, considering the use of fibers for active markers, we characterize the throughput as the longest distance between the end of a fiber and a camera where the emitted blinking pattern of IR light can be detected.
%Therefore, we evaluate the light throughput as the longest distance $d_{cam}$.
According to the inverse square law of illuminance, the longest distance $d_{cam}$ can be expressed as:
\begin{eqnarray}\label{eq:fiber:1}
	d_{cam}=c_4(I_{LED}T_f)^{\frac{1}{2}}=c_5\exp(-\frac{1}{2}\int_fa),
\end{eqnarray}
where $c_4$ and $I_{LED}$ represent an arbitrary coefficient and the light intensity of an IR LED attached to the fiber, respectively, and $c_5=c_4(I_{LED})^{\frac{1}{2}}$.
Once we obtain the parameters $c_1$, $c_2$, $c_3$, and $c_5$, we can compute the light throughput of the fiber in any given route.

%We evaluated the ability of a printed optical fiber in terms of the light throughput.
%We printed out fibers with different lengths and curvatures.
%For each fiber, we checked the longest distance where a blinking pattern of IR light can be identified by a camera.

\begin{figure}[t]
  \centering
  \includegraphics[width=0.98\hsize]{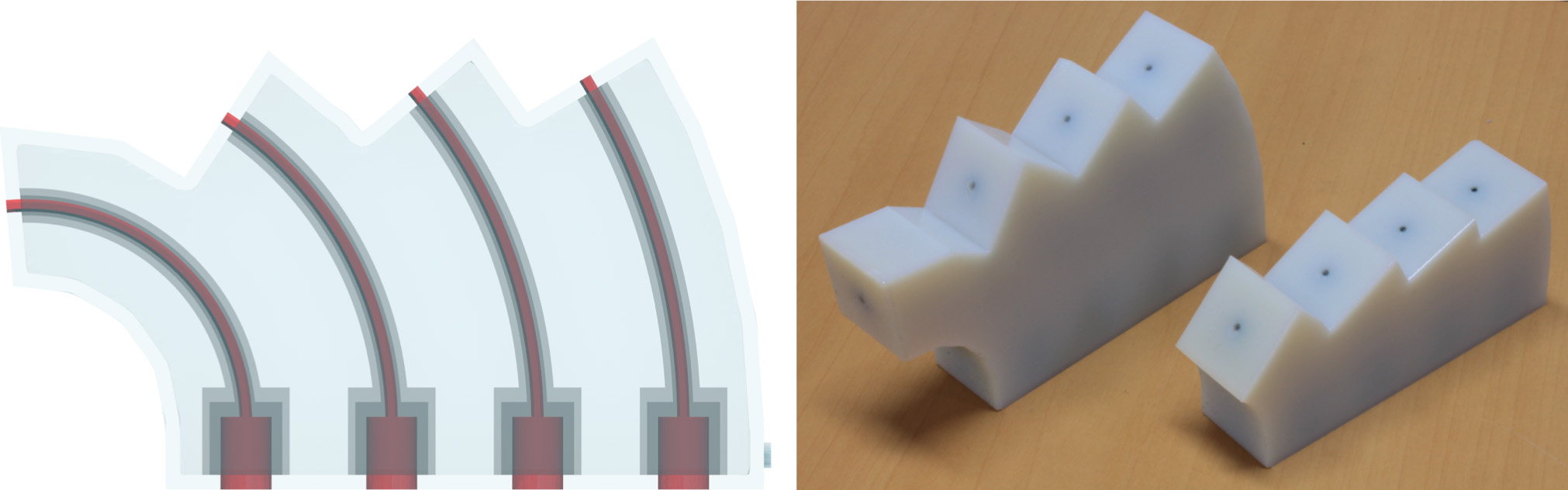}
  \caption{Printed optical fibers for parameter identification: (left) internal structure, (right) printed fibers.}
  \label{fig:light_thruput}
\end{figure}

\begin{figure}[t]
  \centering
  \includegraphics[width=0.98\hsize]{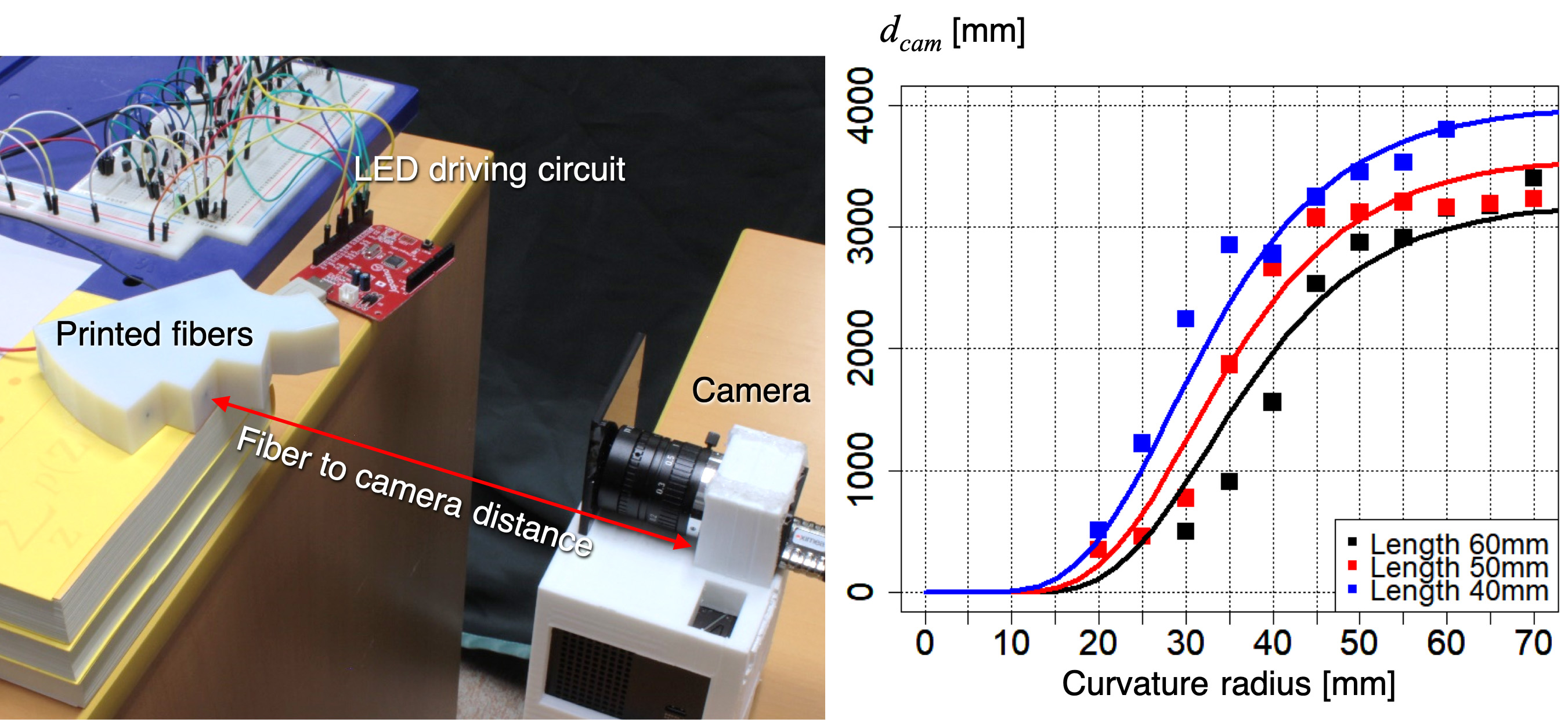}
  \caption{Measurement of $d_{cam}$ (the longest distance between a fiber and camera where blinking binary pattern is detectable): (left) measurement setup, (right) result.}
  \label{fig:exp_thruput}
\end{figure}

To calibrate the parameters, we printed out 33 optical fibers by combining 3 lengths (40, 50, and 60 mm) and 11 curvatures (20, 25, \ldots, and 70 mm) (Figure \ref{fig:light_thruput}).
We attached an IR LED (OSI3CA5111A, 850 nm) onto one end of each fiber and turned it on and off repeatedly.
A camera (XIMEA MQ003MG-CM) captured the LED's blinking pattern emitted from the other end of the fiber.
We changed the distance between the camera and the fiber and recorded the longest distance where the binary pattern can be detected (Fiure \ref{fig:exp_thruput}(left)).
Figure \ref{fig:exp_thruput}(right) shows the results of the experiment.
We fitted the model (Equation (\ref{eq:fiber:1})) to our data (outliers excluded), and obtained the following results: $c_1=2.214\times 10^{-2}$, $c_2=7.478\times 10^{-1}$, $c_3=9.584\times 10^{-2}$, and $c_5=6.224\times 10^3$.
The residual standard error of the fitting was 281.2 mm.

\subsection{Initial routing of printed optical fibers}
\label{sec:fiber:initial}

We compute the routes of optical fibers connecting each IR LED to the corresponding marker points on the projection surface that share the same pattern ID.
The luminance of IR light emitted from the marker points should be maximized so that a camera can detect the blinking patterns from as far distance as possible.
At the same time, collisions between fibers transmitting different pattern IDs must be avoided, and all fibers must be inside the projection object.
In this study, the route is represented as NURBS (Non-Uniform Rational B-Spline) curve.
Our route determination algorithm consists of two parts: initial route computation and refinement.
The former is discussed here, and the latter is discussed in Section \ref{sec:fiber:route}.
%In general, gradient decent algorithms such as Adam are significantly affected by initial solutions.
%For example, when we simply apply straight lines connecting between IR LEDs and corresponding marker points as an initial solution, the gradient of $r_{p_f}$ in Equation \ref{eq:afib} diverged.
%In addition, because $E_{surf}(f)$ evaluates the distance between a fiber and the surface of a projection object, it can be minimized even when the fiber is outside of the object.
%Therefore, the initial route of the fiber must be non-straight and be inside the object.

The initial route should balance the following four demands: (1) the route should be as short as possible and (2) the curvature radius of each part of the fiber should be as large as possible to increase the light throughput, (3) collisions among fibers transmitting different pattern IDs should be avoided, and (4) the fibers should be inside the object.
%In addition, we compute the initial route by considering that a gradient decent algorithm can be applied to it in the refinement phase.

\begin{figure}[t]
  \centering
  \includegraphics[width=0.98\hsize]{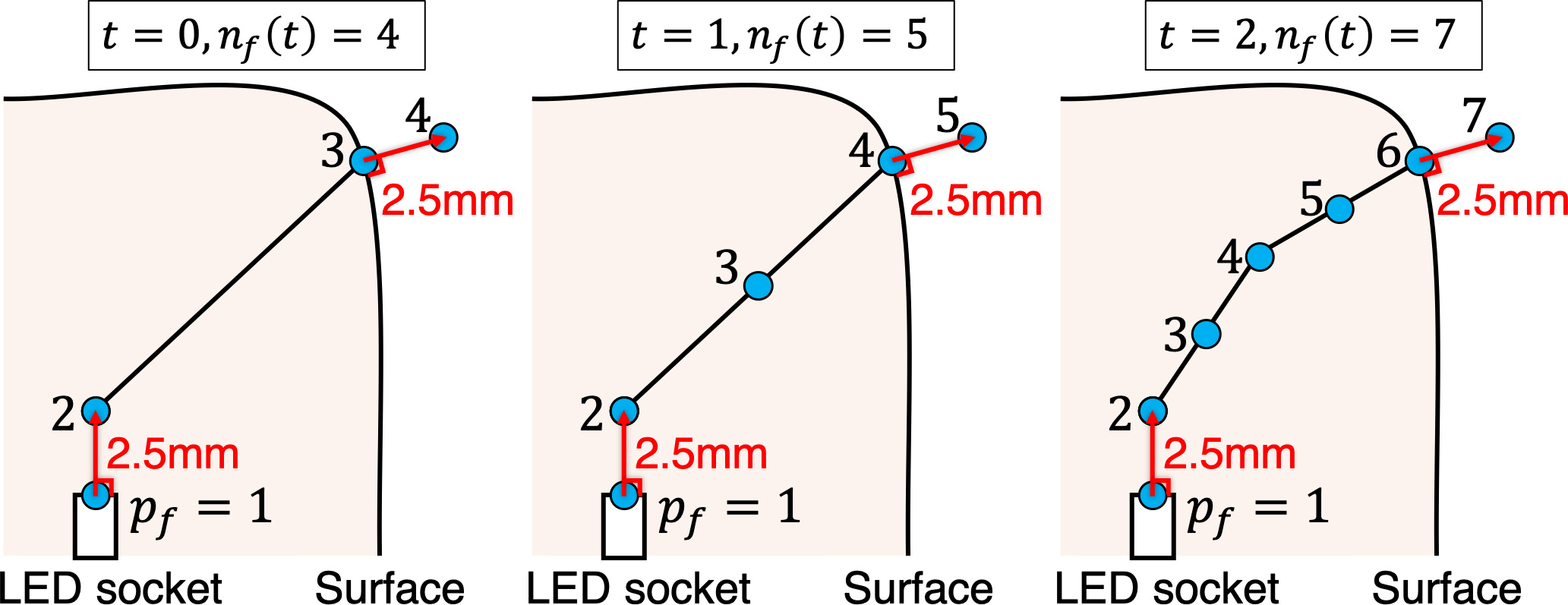}
  \caption{Initial routing of a printed optical fiber.}
  \label{fig:initial_routing}
\end{figure}

At first, we set the initial route of each fiber $f$, which consists of four initial control points ($p_f=1,2,3,4$) as shown in Figure \ref{fig:initial_routing}.
Control points of $p_f=1$ and $p_f=2$ are set at the tip of an IR LED and at 2.5 mm above the LED, respectively.
Those of $p_f=3$ and $p_f=4$ are set at a marker point on the object surface and at 2.5 mm away from it along the normal vector of the point.
Note that these four control points are fixed throughout the initial route computation process. % to maximize the light throughput considering narrow light directivity of IR LED and optical fiber.

Then, we iterate the following two processes: (1) adding new control points to the route and (2) moving all the points except the four initial points.
We denote the number of control points at the $t$-th iteration as $n_f(t)$.
In the former process, we add new control points at every middle position of two adjacent control points excluding the pairs of the points of $p_f=1$ and $2$, and $p_f=n_f(t)-1$ and $n_f(t)$.
Afterward, we update the IDs of control points such that the IDs are assigned in the order of control points from the point at the tip of the IR LED (i.e., $p_f=1$) along the fiber.
%For example, in the first loop of the former process, we add a point of $i_f=4$ at the middle position between $i_f=1$ and $i_f=2$.
%In the second loop, we add two points between $i_f=1$ and $i_f=4$, and between $i_f=4$ and $i_f=2$, and update the IDs such that $i_f=5$ is assigned to a point whose ID was $i_f=4$ in the previous loop.
In the latter process, we slightly move every control point other than $p_f=1, 2, n_f(t)-1, n_f(t)$ according to the following expression:
\begin{equation}\label{eq:heuristic}
	{\bf x}_{p_f}(t+1)={\bf x}_{p_f}(t)+{\bf l}_{p_f}(t)+{\bf r}_{p_f}(t)+{\bf f}_{p_f}(t)+{\bf s}_{p_f}(t),
\end{equation}
where ${\bf x}_{p_f}(t)$ denotes the position of the control point $p_f$ after the $t$-th iteration.

${\bf l}_{p_f}(t)$ adjusts the position ${\bf x}_{p_f}(t)$ such that the fiber becomes shorter.
The shortest length between the points of $p_f=2$ and $n_f(t)-1$ is their Euclidean distance.
We define the division of the shortest length by $(n_f(t)-1)-2$ as the sectional shortest length. %the number of sections between adjacent control points as the sectional shortest length.
${\bf l}_{p_f}(t)$ moves the control point $p_f$ to a direction in which both the distance from ${\bf x}_{p_f-1}(t)$ to ${\bf x}_{p_f}(t)$ and that from ${\bf x}_{p_f}(t)$ to ${\bf x}_{p_f+1}(t)$ get closer to the sectional shortest length.
%When control points of $i_f=1$ and $i_f=2$ are linearly connected, the fiber length between these points becomes the shortest.
%In this case, the averaged length between two adjacent control points is $l_f(t)=|{\bf x}_{i_f=1}-{\bf x}_{i_f=2}|/(n_c(t)+1)$, where $n_c(t)$ is the number of control points excluding those of $i_f=0, 1, 2, 3$.
%When a distance from a control point to an adjacent one becomes very long, the curvature at the point computed based on circumcircle becomes significantly different from the correct value of the NURBS curve.
%This leads to significantly inaccurate absorption computation in Equation \ref{eq:afib}, which will be used in the gradient ascent method (cf. Section \ref{sec:fiber:gam}).
%Therefore, we set the target length between any pair of two adjacent control points as $l_f(t)$, thus:
%
%\begin{eqnarray}
%	{\bf l}_{i_f}(t)=(|{\bf x}_{i_f-1}(t)-{\bf x}_{i_f}(t)|-l_f(t))\frac{{\bf x}_{i_f-1}(t)-{\bf x}_{i_f}(t)}{|{\bf x}_{i_f-1}(t)-{\bf x}_{i_f}(t)|} \nonumber \\
%	+(|{\bf x}_{i_f+1}(t)-{\bf x}_{i_f}(t)|-l_f(t))\frac{{\bf x}_{i_f+1}(t)-{\bf x}_{i_f}(t)}{|{\bf x}_{i_f+1}(t)-{\bf x}_{i_f}(t)|}.
%\end{eqnarray}
%
${\bf r}_{p_f}(t)$ adjusts the position ${\bf x}_{p_f}(t)$ so that the curvature radius of the fiber around the control point $p_f$ becomes larger.
It moves the control point away from the center of the circle passing through ${\bf x}_{p_f-1}(t)$, ${\bf x}_{p_f}(t)$, and ${\bf x}_{p_f+1}(t)$.
%We define it as follows:
%
%\begin{equation}
%	{\bf r}_{i_f}(t)=\{l_f(t)\}^2(\frac{{\bf x}_{i_f}(t)-{\bf c}_{i_f-1}(t)}{|{\bf x}_{i_f}(t)-{\bf c}_{i_f-1}(t)|^2}+\frac{{\bf x}_{i_f}(t)-{\bf c}_{i_f+1}(t)}{|{\bf x}_{i_f}(t)-{\bf c}_{i_f+1}(t)|^2}).
%\end{equation}
%
${\bf f}_{p_f}(t)$ adjusts the position ${\bf x}_{p_f}(t)$ to avoid collision of the fiber $f$ and another fiber $f'$ transmitting a different pattern ID.
Suppose there are multiple control points of $f'$ that are closer to the control point $p_f$ than a predefined distance, we denote them as $\hat{p}_{f'}$.
Then, ${\bf f}_{p_f}(t)$ denotes the sum of vectors from each ${\bf x}_{\hat{p}_{f'}}$ to ${\bf x}_{p_f}$.
%We compute it by adding the unit vectors from each of other control points $i_{f'}$ to the current point $i_f$ only if the distance between these two points are less than a threshold $d_{th}$, thus;
%
%\begin{equation}
%	{\bf a}_{i_f}(t)=\sum_{\{f'\mid f'\ne f\}}\sum_{\{i_{f'}\mid d_{th}>|{\bf x}_{i_f}-{\bf x}_{i_{f'}}|\}}\frac{{\bf x}_{i_f}(t)-{\bf x}_{i_{f'}}(t)}{|{\bf x}_{i_f}(t)-{\bf x}_{i_{f'}}(t)|}.
%\end{equation}
%
${\bf s}_{p_f}(t)$ moves the control point $p_f$ away from the closest surface point when the distance between these two points is shorter than a predefined threshold.

In each iteration, we operate the former process (i.e., adding new control points) once, and then the latter process (i.e., moving them) 1,000 times.
After five iterations, we obtained 35 control points in total whose positions are the initial route of each printed optical fiber.

\subsection{Route optimization}
\label{sec:fiber:route}

We refine the initial route using the following optimization framework.
The target of the route optimization is to find the optimal set of $f$ such that
\begin{equation}\label{eq:opt}
	{\rm minimize}\ \sum_f [\{1-E_{cam}(f)\}+k_3E_{fib}(f)+k_4E_{surf}(f)+k_5E_{reg}(f)],
\end{equation}
where $E_{cam}(f)$, $E_{fib}(f)$, and $E_{surf}(f)$ are used to evaluate the light throughput of the fiber $f$, the collision between $f$ and other fibers, and the distance of $f$ from the object surface, respectively.
$E_{reg}(f)$ works as a regularizer.
$k_3$, $k_4$, and $k_5$ are arbitrary weights.
The optimization is performed using Adam (Adaptive Moment Estimation) algorithm \cite{kingma2014adam}.

$E_{cam}(f)$ represents the estimated maximum distance from the marker point of $f$ to a place where the camera used in Section \ref{sec:fiber:throughput} can detect the blinking pattern.
%We assume the route of a fiber is a combination of curves with different lengths and curvatures.
We approximate the route of fiber as consisting of arcs belonging to different control points.
The arc of a control point $p_f$ is defined as follows.
First, it is a part of the circumscribed circle passing through ${\bf x}_{p_f-1}$, ${\bf x}_{p_f}$, and ${\bf x}_{p_f+1}$.
Second, its length $l_{p_f}$ is half of the sum of the distances from ${\bf x}_{p_f-1}$ to ${\bf x}_{p_f}$ and that from ${\bf x}_{p_f}$ and ${\bf x}_{p_f+1}$.
%; thus, we cannot directly use $d_{cam}$.
Due to the multiplicative nature of light absorption, we model $E_{cam}(f)$ as the product of $d_{cam}$ at each control point based on Equation (\ref{eq:fiber:1}) as
\begin{equation}\label{eq:ecam}
	E_{cam}(f)=\prod_{p_f} \exp(-\frac{1}{2}l_{p_f}a_{p_f}),
\end{equation}
where
\begin{eqnarray}
	l_{p_f}=\frac{1}{2}(|{\bf x}_{p_f-1}-{\bf x}_{p_f}|+|{\bf x}_{p_f+1}-{\bf x}_{p_f}|), \label{eq:lfib} \\
	a_{p_f}=c_1+c_2\exp(-c_3r_{p_f}), \label{eq:afib}
\end{eqnarray}
and $r_{p_f}$ denotes the radius of the circumscribed circle.
$E_{fib}(f)$ increases when the distance from the fiber $f$ to other fibers $f'$ (through which different blinking patterns are transmitted), decreases.
This term helps to prevent crosstalk in the blinking patterns.
%Specifically, we compute the sum of the distance from each control point of $f$ to all the control points of other fibers $f'$ which meet the condition of $u_{ptn}(f)\ne u_{ptn}(f')$, where $u_{ptn}(f)$ represents the pattern ID of IR light transmitted through the fiber $f$.
%We compute the sum of the distances from all the control points of $f$ to those of all the fibers $f'$.
Thus,
\begin{equation}\label{eq:fiber:2}
	E_{fib}(f)=\sum_{p_f}\sum_{f'}\sum_{p_{f'}}\exp(-k_6|{\bf x}_{p_f}-{\bf x}_{p_{f'}}|),
\end{equation}
where $k_6$ represents an arbitrary coefficient.
$E_{surf}(f)$ increases when the distance between the fiber $f$ and the surface of the projection object decreases.
This term constrains the fiber to remain inside the projection object.
Suppose a surface point of the projection object and its position are denoted as $p_s$ and ${\bf x}_{p_s}$, respectively, $E_{surf}(f)$ can be computed using the following expression: %in the manner as Equation (\ref{eq:fiber:2}).
\begin{equation}
	E_{surf}(f)=\sum_{p_f}\sum_{p_s}\exp(-k_7|{\bf x}_{p_f}-{\bf x}_{p_s}|),
\end{equation}
where $k_7$ represents an arbitrary coefficient.
The last term, $E_{reg}(f)$, plays the role of a regularizer.
For each control point $p_f$, if the distance from the previous control point $p_f-1$ is significantly different from that from the next control point $p_f+1$, the approximation of the route as a group of arcs in Equation (\ref{eq:ecam}) will no longer be valid.
Thus, we formulate $E_{reg}(f)$ to minimize the difference between the two distances as:
\begin{equation}
	E_{reg}(f) = \sum_{p_f}\left||{\bf x}_{p_f-1}-{\bf x}_{p_f}|-|{\bf x}_{p_f+1}-{\bf x}_{p_f}|\right|_2.
\end{equation}
%

%%%%%%%%%%%%%%%%%%%%%%%%%%%%%%%%%%%%%%%%%%%%%%%%%%
\section{Experiments}
%%%%%%%%%%%%%%%%%%%%%%%%%%%%%%%%%%%%%%%%%%%%%%%%%%

We fabricated three objects with optical fibers and conducted experiments to validate the effectiveness of the proposed method.
First, we introduce the details of the fabricated objects and a prototype PM system.
Second, we present an evaluation experiment of pose estimation errors.
Finally, we present dynamic PM results.

\subsection{Printed objects and experimental system}

\begin{figure}[t]
  \includegraphics[width=0.98\hsize]{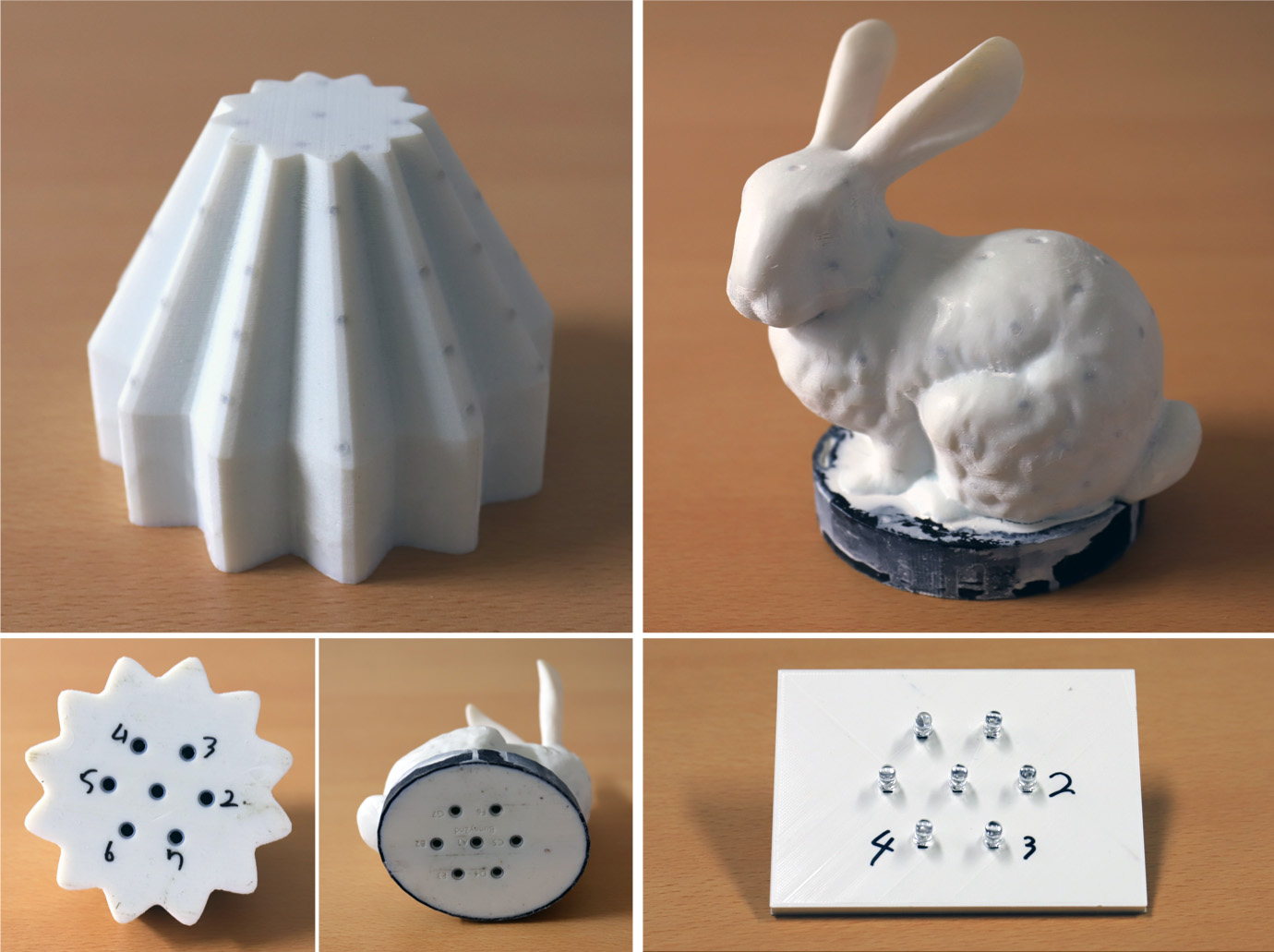}
  \caption{Printed objects and IR LEDs of the base component (bottom right).}
  \label{fig:printed}
\end{figure}

We printed out three target objects with different shapes (wavy-cone, building, and bunny) using Stratasys Objet260 Connex3 as shown in Figures \ref{fig:teaser} and \ref{fig:printed}.
We modeled a wavy-cone shape to show that the proposed method works even for a surface having symmetric structure and strongly uneven and curved shape, for which conventional marker-less and spatial-pattern marker-based methods theoretically do not work well as discussed in Section \ref{sec:relatedworks}.
As a more practical projection target, we modeled the building surface by assuming to use it in an appearance design scenario.
It also has an almost symmetric structure and some uneven parts.
We selected the Stanford bunny to show the robustness of our method against occlusions (i.e., the ears occlude the body when viewed from a certain viewing area).
The sizes of the objects were $111\times111\times87$ [mm] (wavy-cone), $87\times152\times101$ [mm] (building), and $110\times138\times148$ [mm] (bunny).

\begin{table}[t]
  \centering
  \caption{Binary codes for seven LEDs.}
  \begin{tabular}{l|l}
    M-sequence & 100011110101100 \\
    \hline
    Others & 001, 010, 011, 100, 101, 110
  \end{tabular}
  \label{tab:codes}
\end{table}

We prepared seven small holes in the bottom surfaces of the objects as LED sockets as shown in Figures \ref{fig:teaser} and \ref{fig:printed}.
The same hole layout was shared among the objects so that the same base component consisting of an electrical circuit, battery, and LEDs is reusable.
The center of the holes was used for an LED transmitting an m-sequence code, and the others were used for the other binary codes.
Therefore, the number of codes $N_{ptn}$ was 7 (Table \ref{tab:codes}).
We used an m-sequence whose code length was 15 (i.e., $m=4$ in Section \ref{sec:marker:pattern}).
We set the code length of the other binary codes $b_{ptn}$ as 3.
We determined the marker placements for all the objects using our method in Section \ref{sec:marker:place} where the parameters $k_1$ and $k_2$ were set as 5 and 2, respectively.
As a result, 43 (wavy-cone), 57 (building), and 65 (bunny) markers were placed on the objects' surfaces.

\begin{figure}[t]
  \includegraphics[width=0.98\hsize]{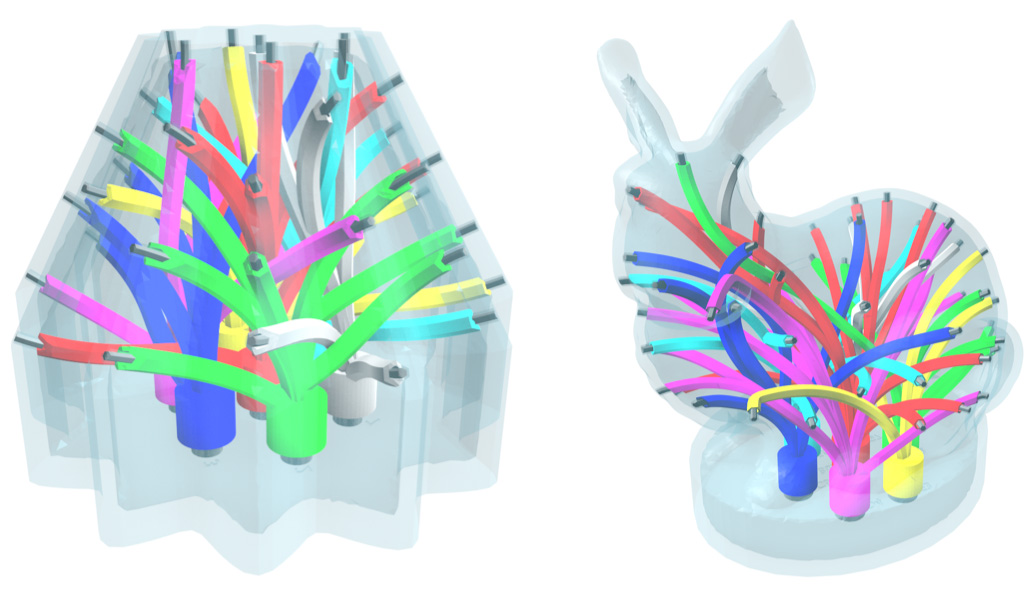}
  \caption{The internal structures of the printed objects. Fibers with the same color are connected to the same LED.}
  \label{fig:internal_structure}
\end{figure}

The routes of optical fibers were optimized using the proposed method as discussed in Sections \ref{sec:fiber:initial} and \ref{sec:fiber:route}.
The parameters of $k_3$, $k_4$, $k_5$, $k_6$, and $k_7$ were set as 1.5, 0.5, 1.0, 1.5, and 1.5, respectively.
The internal structures of the objects were shown in Figures \ref{fig:teaser} and \ref{fig:internal_structure}.
To evaluate the effectiveness of the optimization method, we checked $E_{cam}(f)$, the computed longest distance from each marker point to the camera used in Section \ref{sec:fiber:throughput}.
The average value of $E_{cam}(f)$ of the initial route was 1618.7 mm and that of the optimized route was 2028.2 mm in the wavy-cone object.
The average values of $E_{cam}(f)$ were improved in the other objects, too (building: 868.4 mm to 1407.9 mm, bunny: 843.0 mm to 1543.0 mm).
In addition, the fibers of different pattern IDs did not collide with each other, and all the fibers were placed inside the objects.
Particularly, in the bunny object, fibers were successfully routed through the narrow neck.

\begin{figure}[t]
  \includegraphics[width=0.98\hsize]{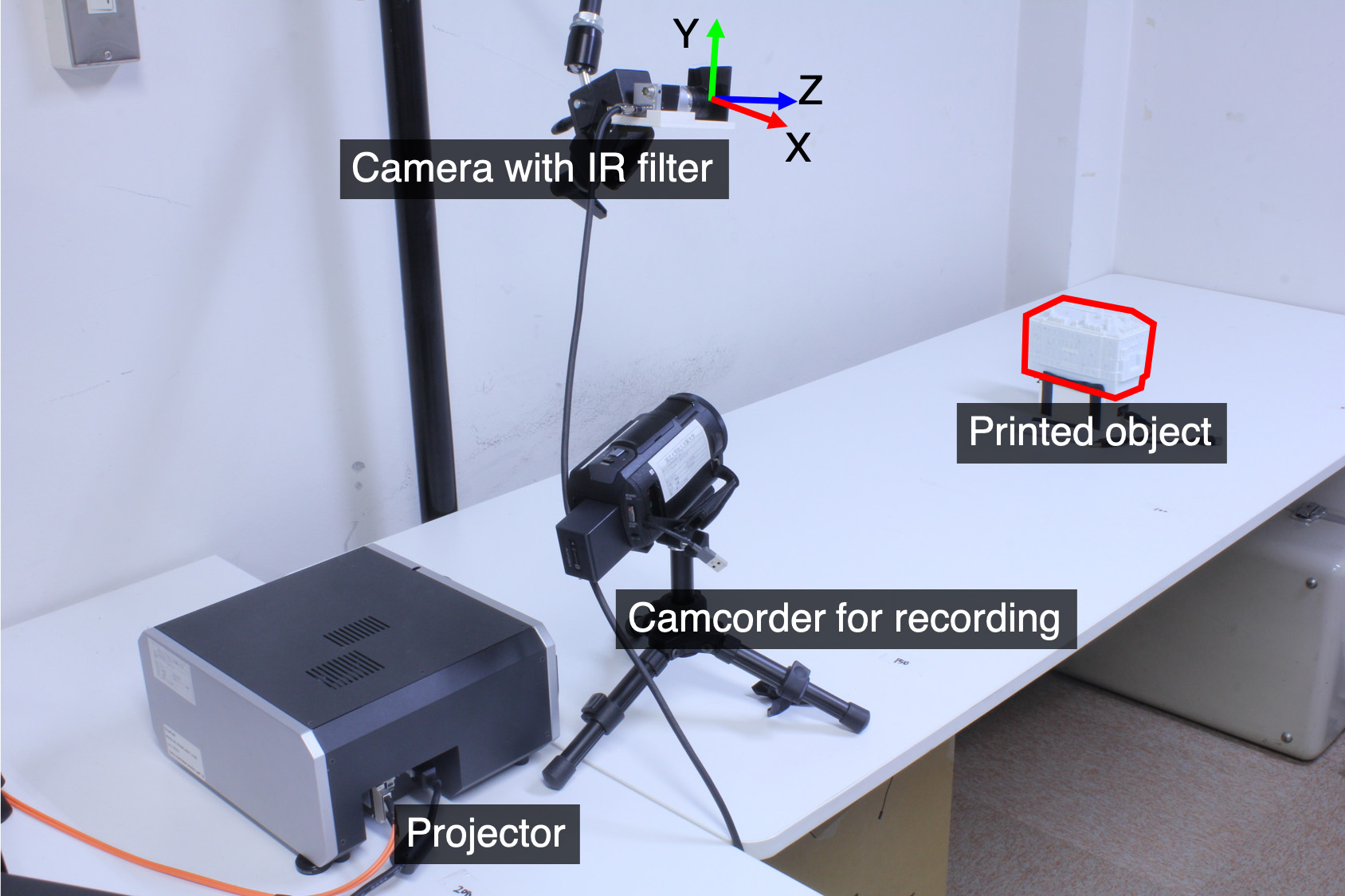}
  \caption{Projector-camera system. The colored axes show the IR camera coordinate system.}
  \label{fig:procams}
\end{figure}

%\begin{figure}[t]
%  \includegraphics[width=0.98\hsize]{figure/base_component.jpg}
  %\caption{Key devices of the base component (from left to right: microcoputer, IR LEDs, and mobile battery).}
%  \caption{IR LEDs of the base component.}
%  \label{fig:base_component}
%\end{figure}

We built a projector-camera system as shown in Figure \ref{fig:procams}.
We applied an off-the-shelf industrial camera (Basler acA720-520um, 525 fps, 720$\times$540 pixels) with an IR-pass/VIS-cut filter.
It has been pointed out that low-latency augmentation is crucial in dynamic PM applications \cite{bermano2017makeup,narita2017dynamic,watanabe2017extended,Miyashita:2018:MPM:3272127.3275045}.
Therefore, we applied a 1,000 fps projector (Inrevium, TB-UK-DYNAFLASH, 8-bit grayscale, 1024$\times$768 pixels).
The camera and projector were connected to a PC (CPU: Intel Core-i7 5960X 3.0 GHz, RAM: 32 GB).
The base component consisted of seven LEDs (OSI3CA5111A, 850 nm), an Arduino-compatible microcomputer (Japanino) for controlling the LEDs, and a mobile battery (Figure \ref{fig:printed}).
\revise{}{The blinking speed was 350 bit/s.}
A camcorder in the figure was used to record videos of projected results.

\subsection{Evaluation of pose estimation}
\label{subsec:exp_estimation}

\begin{figure}[t]
  \includegraphics[width=0.98\hsize]{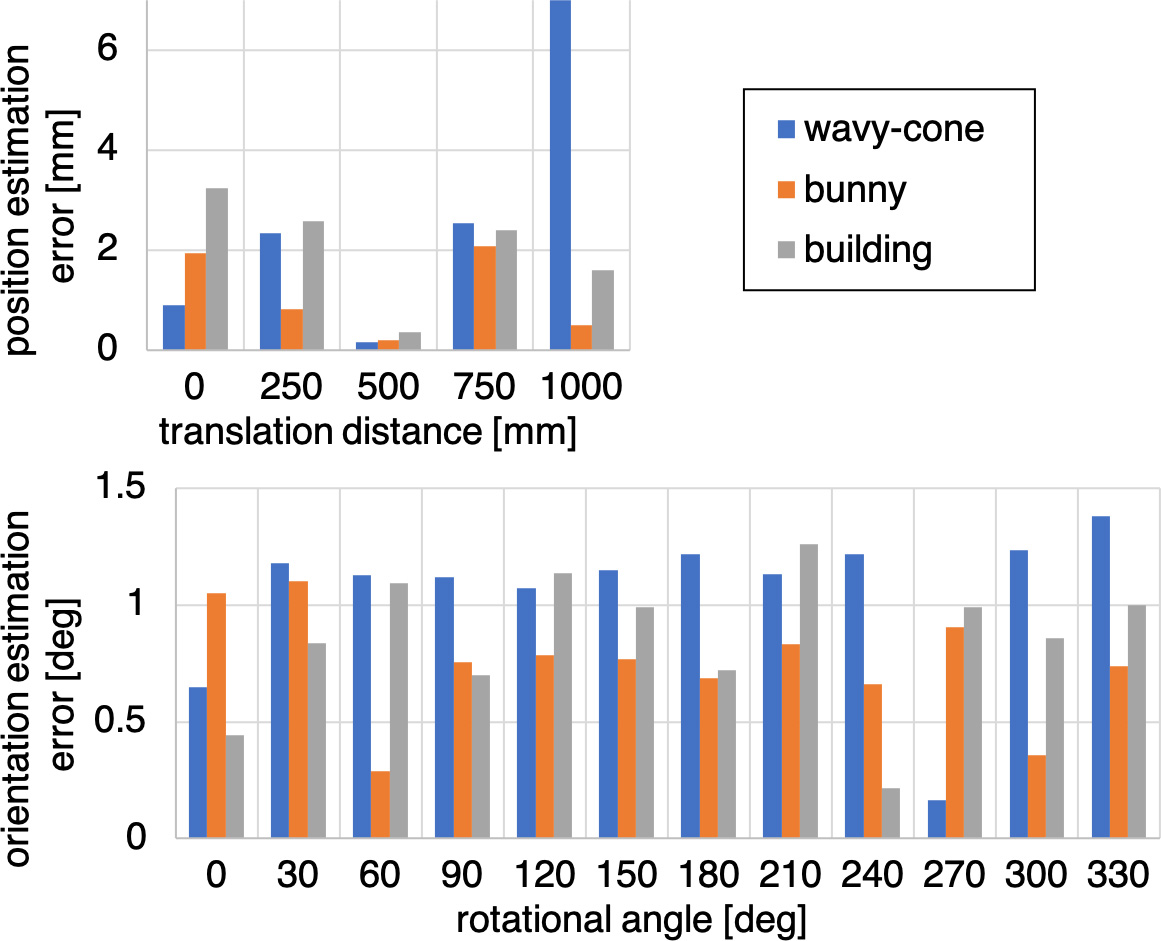}
  \caption{Pose estimation error: (\revise{left}{top}) position estimation error for a translation, (\revise{right}{bottom}) orientation estimation error for a rotation.}
  \label{fig:error}
\end{figure}

We quantitatively evaluated the pose estimation accuracy of the proposed method by measuring two errors; one for translation and the other for rotation.
First, we translated each of the objects (wavy-cone, bunny, and building) along a straight line, and estimated its position in the IR camera coordinate system at five locations.
The distance between the adjacent locations is 250 mm, and thus, the overall measurement range was 1 m.
Figure \ref{fig:error}(\revise{left}{top}) shows the Euclidean distance between the grand truth and the estimated position.
We confirmed that the errors were 1.9 mm on average and less than 7.0 mm at any estimated locations in the range of 1 m.

Second, we rotated each of the objects about an axis using a turntable, and estimated its pose every 30 degrees.
Figure \ref{fig:error}(\revise{right}{bottom}) shows the absolute errors of the estimated rotational angles about the axis.
We confirmed that the errors were 0.9 degrees on average and less than 1.4 degrees at any rotational angle.
%In summary of the two evaluations, considering the size of the objects (approximately $xx\times xx\times xx$ [mm]), we believe that the errors were sufficiently small to geometrically align projection images with perceptually acceptable accuracies.

% 距離の範囲が広いこと。
% wavy coneでもトラッキング出来ていること

\subsection{Dynamic projection mapping experiment}
\label{subsec:exp_projection}

We conducted a dynamic PM experiment using the bunny and building objects.
We moved the objects by hands in front of the projector-camera system.
The movement consisted of translations along the xyz-axes of the IR camera coordinate system and rotation about the y-axis.
We also covered the objects with hands to investigate the robustness against occlusion.
We tried to hold the objects such that they could not move while covering them.

\begin{figure*}[t]
  \includegraphics[width=0.59\hsize]{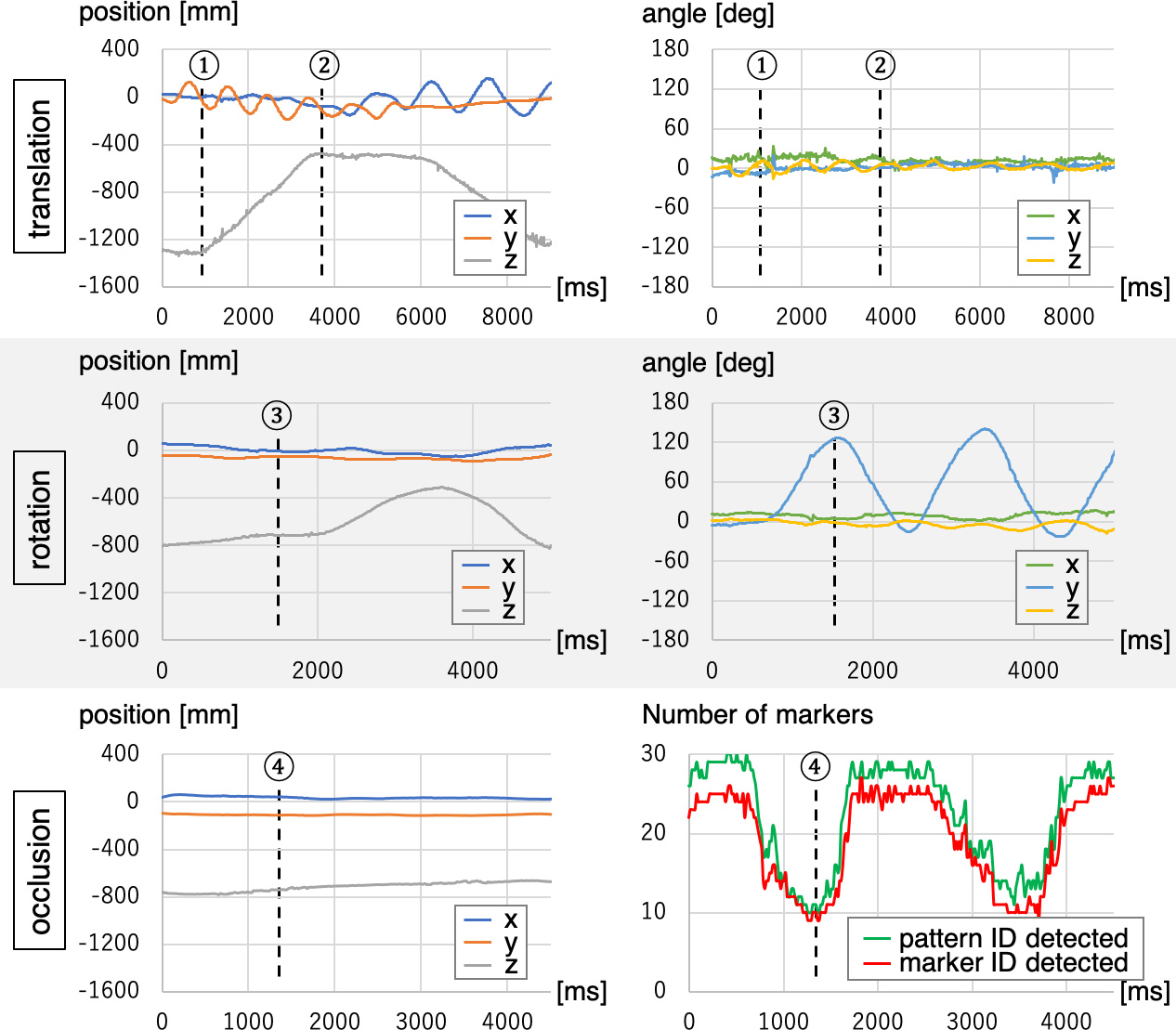}
  \includegraphics[width=0.4\hsize]{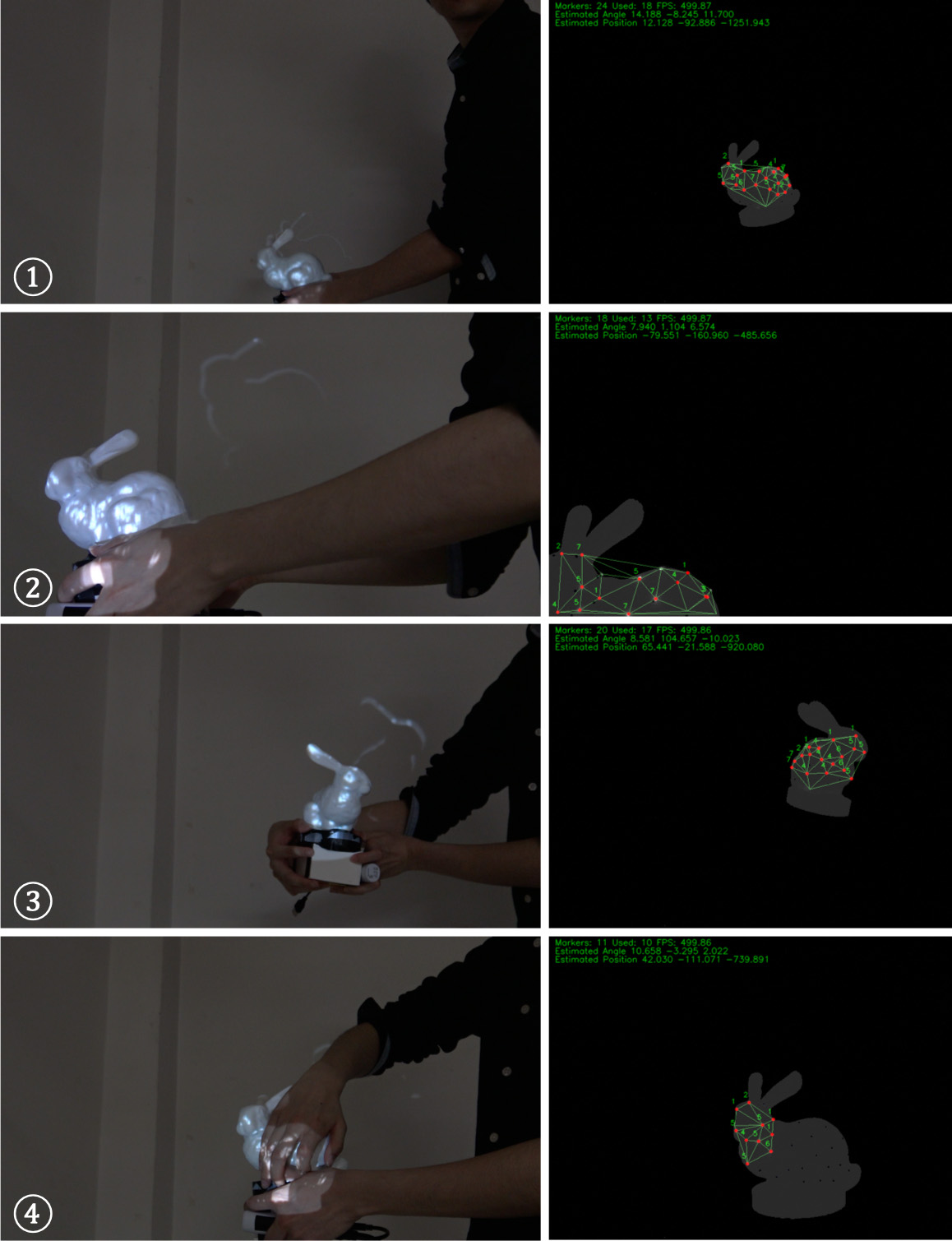}
  \caption{Projection result of the bunny object in translation, rotation, and occlusion: (left) the time series of 3D position and orientation of the object in the IR camera coordinate system, and the number of markers whose pattern/marker IDs were detected, (right) captured image of the projected object by the camcorder and visualization of detected markers and estimated pose of the object superimposed in captured image by the camera at frames indicated by the numbers in a circle.}
  \label{fig:res_bunny}
\end{figure*}

\begin{figure*}[t]
  \includegraphics[width=0.59\hsize]{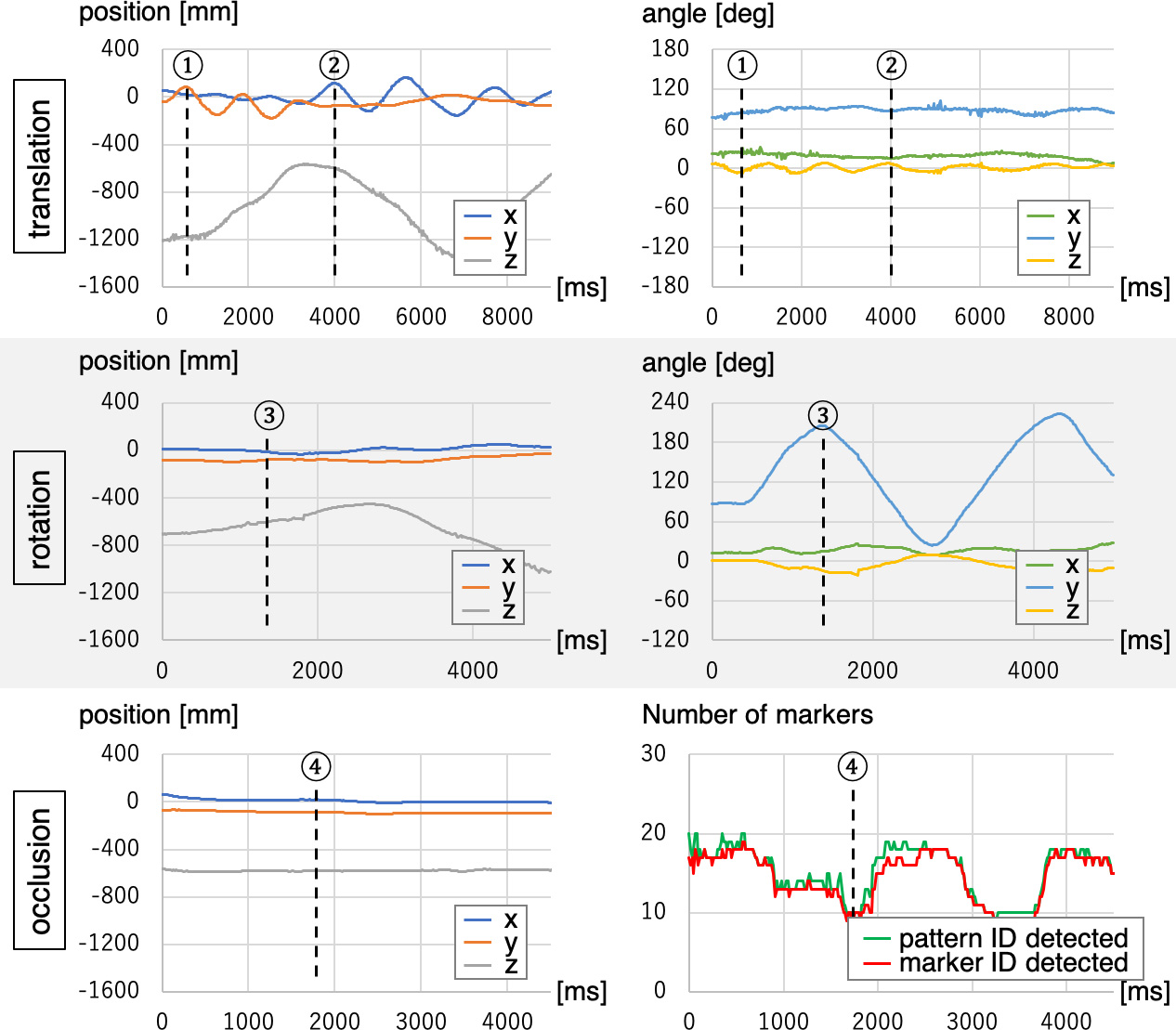}
  \includegraphics[width=0.4\hsize]{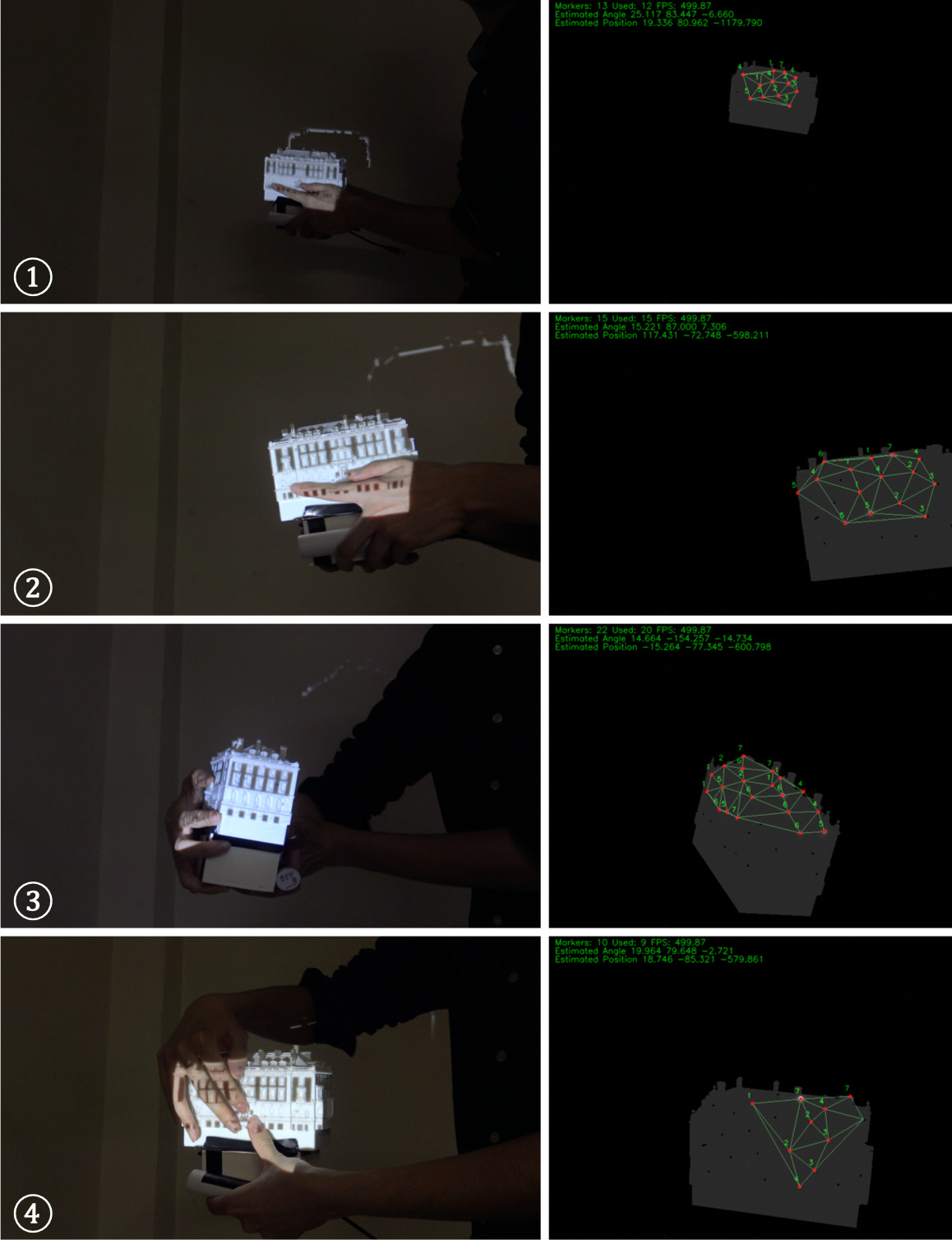}
  \caption{Projection result of the building object in translation, rotation, and occlusion: (left) the time series of 3D position and orientation of the object in the IR camera coordinate system, and the number of markers whose pattern/marker IDs were detected, (right) captured image of the projected object by the camcorder and visualization of detected markers and estimated pose of the object superimposed in captured image by the camera at frames indicated by the numbers in a circle.}
  \label{fig:res_building}
\end{figure*}

Figures \ref{fig:res_bunny} and \ref{fig:res_building} show the results of the bunny and building objects, respectively.
From the results of translation and rotation, we confirmed that the projection images could be aligned onto both objects which were largely translated (approx. 800 mm along z-axis) and rotated (approx. 180 degrees about y-axis).
From the results of occlusion experiments conducted, we confirmed that the projected images were stably aligned onto both objects even when more than half of the object was occluded by the hands.
Although the numbers of detected markers were decreased owing to the occlusions, no significant error was observed in the pose estimation of the objects.
\revise{}{The processing time for a captured image was 2-3 ms.}
In addition, we \revise{}{informally} confirmed that the markers were \revise{not noticeable}{unnoticeable} to human observers under projection.

%%%%%%%%%%%%%%%%%%%%%%%%%%%%%%%%%%%%%%%%%%%%%%%%%%
\section{Discussion}
\label{sec:discussion}
%%%%%%%%%%%%%%%%%%%%%%%%%%%%%%%%%%%%%%%%%%%%%%%%%%

From the results of the pose estimation evaluation presented in Section \ref{subsec:exp_estimation}, we consider the pose estimation errors (1.9 mm in position and 0.9 degrees in orientation estimations) sufficiently small to geometrically align projection images with perceptually acceptable accuracies, considering the size of the objects (approximately $120\times120\times120$ [mm]).
\revise{}{There was a relatively large error ($>$ 6 mm) in the wavy-cone position estimation at 1 m. This might happen due to that the size of the object is smaller and consequently the markers are placed closer to each other than the other objects.}
The results of the PM experiment in Section \ref{subsec:exp_projection} show that robust and accurate dynamic PM was achieved with the objects moving and rotating in a large space and even while they were occluded.
This result indicates that the proposed algorithm successfully determined the marker places and fiber routes for various shapes.

We confirmed two major advantages of the fabricated optical fiber-based markers over previous visual markers for dynamic PM \cite{asayama2018fabricating,watanabe2017extended}.
First, our method can be small enough to be embedded on strongly uneven or curved surfaces such as the wavy-cone and building surfaces, for which conventional visual markers theoretically do not work due to self-occlusion and deformations.
Therefore, the proposed markers can relax the planarity constraint of a target surface shape.
Second, the results of the pose estimation evaluation experiment showed that the working range of the proposed method was more than 1,000 mm.
This was significantly larger than the working ranges reported in previous works (150 mm \cite{asayama2018fabricating} and 350 mm \cite{watanabe2017extended}).

\revise{}{Camera-based motion capture system typically applies multiple cameras and measures the 3D positions of markers by stereo. The measured 3D point set can solve the pose estimation of a rigid-body projection object using the Iterative Closest Point algorithm, if the marker locations on the object are known in advance. This means that we do not need marker identifications, and consequently, the marker design can be simpler in a multi-camera system. On the other hand, the single camera approach applied in this research works without any well-configured motion capture systems. We also consider to extend our method to track non-rigid surfaces, which will be discussed in the next subsection.}

\subsection{\revise{}{Limitation}}

\revise{}{We discuss limitations of the proposal. First, the current system applies the fixed shape base component. Although it is convenient and useful, it restricts the shape of a projection object. To overcome this limitation, our system needs to allow the electronic devices of the base component (e.g., LEDs and a battery) to be positioned within the projection object of an arbitrary shape. To this end, we need to extend our current optimization framework such that not only optical fiber routes but the placements of the electronic devices are jointly optimized. We consider that this can be simply achieved by adding new constraints about the placements in the current optimization framework.}

\revise{Limitations of the method are mainly}{Second, we discuss limitations} due to the current 3D printing technology.
We used a state-of-the-art multi-material 3D printer.
However, because the current clear material is not perfectly transparent, which lowers the light throughput of a fabricated fiber, both the length and curvature of the fiber would be limited.
As a result, we could not make the projection objects larger than a certain size.
Also, the aperture of our fabricated fiber (core diameter: 1.75 mm) was much larger than the commercially available optical fiber (core diameter: 0.01--0.06 mm) owing to the low spatial resolution of the current printer.
This limits the number of fibers embedded in a projection object, and consequently, the number of markers.
\revise{}{Our method theoretically works for any shapes as long as marker points on the surface and the bottom surface is connected. However, in practice, the above-mentioned large aperture problem limits the shape of a projection object to thick and large shapes. Our method is not applicable to an object consisting of thinner or narrower shapes than the fiber aperture (e.g., hourglass shape).}
Because 3D printing technology is evolving at an accelerated rate, we believe that these limitations will be solved in the future.

Future advancements in 3D printing technology also lead to interesting future research directions.
Specifically, once a printer supports \revise{fabricate}{fabricating} flexible optical fibers, we can achieve novel types of dynamic PMs.
For instance, we can align projected images onto articulated multiple rigid surfaces such as robots.
We can also realize a dynamic PM on a non-rigid, deformable surface using the proposed method.

%%%%%%%%%%%%%%%%%%%%%%%%%%%%%%%%%%%%%%%%%%%%%%%%%%
\section{Conclusion}
%%%%%%%%%%%%%%%%%%%%%%%%%%%%%%%%%%%%%%%%%%%%%%%%%%

This paper presented a novel active marker for dynamic PM, which emits a temporal blinking pattern of IR light representing its ID.
We applied a multi-material 3D printer to fabricate a projection object with optical fibers, which guided IR light from LEDs attached on the bottom of the object.
We proposed an automatic marker placement algorithm to spread multiple active markers over the surface of a projection object in such a way that its pose can be robustly estimated using captured images from arbitrary directions.
We also developed an optimization framework for determining the routes of the optical fibers so that collisions of the fibers were avoided and loss of light intensity in the fibers minimized.
Through experiments conducted, we confirmed that the markers can be embedded on strongly curved surfaces and \revise{may not be perceivable}{informally confirmed that they are unnoticeable} under projection.
In addition, the working range of our system (1 m) was significantly larger than that of previous marker-based methods ($<$350 mm).
We were also able to accurately align projected images onto an object with relatively small pose estimation errors (1.9 mm in position and 0.9 degrees in orientation estimations).
Based on these results, we believe that our proposed method has the potential to expand the applicability of dynamic PM.

\revise{}{We consider the following technical extension. The current system assigns static pattern IDs (LEDs' blinking patterns) to the markers. As future work, we will dynamically assign the IDs only to a part of markers which are visible from the camera. By this, we can reduce the number of the IDs and shorten their bit depths. Consequently, the pose estimation of a fast moving object can be more robust.}

%In addition to the future directions discussed in Section \ref{sec:discussion}, we consider the following two technical extensions.
%First, we will refine the base component to be smaller and thinner than the current one.
%Second, we will try to automatically determine values for the parameters of our marker placement and fiber routing algorithms, which were determined manually in the experiments.

\acknowledgments{
This work has been supported by the JSPS KAKENHI under grant number 15H05925.}

\bibliographystyle{abbrv-doi}

\bibliography{template}
\end{document}